# Phase segregation facilitates exfoliation of franckeite crystals to a single unit cell thickness

Matěj Velický,*[1] Peter S. Toth,[1] Alexander M. Rakowski,[2] Aidan P. Rooney,[2] Aleksey Kozikov,[3] Artem Mishchenko,[3] Colin R. Woods,[3] Thanasis Georgiou,[4] Sarah J. Haigh,[2] Kostya S. Novoselov,[3] and Robert A.W. Dryfe*[1]

[1] School of Chemistry, University of Manchester, Oxford Rd, Manchester, M13 9PL, UK

[2] School of Materials, University of Manchester, Oxford Rd, Manchester, M13 9PL, UK

[3] School of Physics and Astronomy, University of Manchester, Oxford Rd, Manchester, M13 9PL, UK.

[4] Manchester Nanomaterials Ltd, 83 Ducie Street, Manchester, M1 2JQ, UK

* To whom correspondence should be addressed. Tel: +44 (0)161-306-4522; Fax: +44 (0)161-275-4598, e-mail: matej.velicky@manchester.ac.uk or robert.dryfe@manchester.ac.uk.

**Abstract**

Weak interlayer van der Waals interactions in bulk crystals facilitate their mechanical exfoliation to monolayer and few-layer two-dimensional (2D) materials, which exhibit striking physical phenomena absent in their bulk form. Here we study a 2D form of a mineral franckeite and show that phase segregation into discrete layers at the sub-nanometre scale facilitates its layered structure and basal cleavage. This behaviour is likely to be common in a wider family of complex crystals and could be exploited for a single-step synthesis of van der Waals heterostructures, as an alternative to stacking of 2D materials. Mechanical exfoliation allowed us to produce crystals down to a single unit cell thickness and rationalise its basal cleavage by atomic-resolution scanning transmission electron microscopy (STEM). We demonstrate p-type electrical conductivity and remarkable electrochemical properties in exfoliated crystals, which shows promise for energy storage applications.

Research on 2D materials, has so far been mainly focused on unary and binary crystals, such as graphene and $MoS_2$,[1] and little attention has been paid to more complex layered materials.[2] Furthermore, preferential phase segregation, strongly dependent on chemical composition, has been previously observed in ternary sulphides such as $PbSnS_2$.[3] Here, we study a thermodynamically stable, mixed-metal sulphide franckeite, composed of lead, tin, antimony, iron, and sulphur. Franckeite exhibits phase segregation into discrete layers held together by van der Waals forces, which facilitates its basal cleavage, and we show that it can be exfoliated to single unit cell thickness (~2 nm). This results in a high aspect ratio between the lateral size and thickness of the exfoliated crystals, as confirmed using optical microscopy, atomic force microscopy (AFM), and Raman spectroscopy. Scanning electron microscopy (SEM), energy-dispersive X-ray spectroscopy (EDXS), and X-ray photoelectron spectroscopy (XPS) are used to determine franckeite structure and composition. The band gap of the basic metal sulphides, which combine to constitute franckeite ranges from 0.37 eV in PbS (galenite) to 2.1 eV in $SnS_2$ (berndtite),[4] with franckeite itself having an infrared band gap of 0.65 eV.[5] This offers great opportunities in band gap engineering,[6] phase engineering,[7] thermoelectric materials,[3] and solar control coatings,[8] which could be realised through synthesis of complex metal-sulphides with on-demand properties.

The SEM images in Fig. 1a−c show the layered nature of franckeite, which facilitates its mechanical exfoliation. The highest magnification image (Fig. 1c) details terraces of micro-/nano-scopic width, whose lateral dimensions match the typical dimensions of monolayer and few-layer crystals exfoliated onto a $SiO_2$ substrate. The EDXS elemental maps in Fig. 1d−h show that the main elements, lead, tin, antimony, iron, and sulphur, are homogeneously distributed when viewed perpendicular to the layering direction. The averaged EDXS spectrum and quantification in Fig. 1i were the basis for the composition stoichiometry analysis, resulting in an approximate chemical formula of $Pb_6Sn_3Sb_2FeS_{12}$. This matches franckeite, a member of a family of complex metal sulphide minerals including cylindrite, potosíite, and incaite, which are found in the southwest of Bolivia, and have a generic chemical formula $(Pb, Sn)_{6+x}^{2+} Sb_2^{3+} Fe^{2+} Sn_2^{4+} S_{14+x}$, where $-1 \leq x \leq 0.25$.[9]

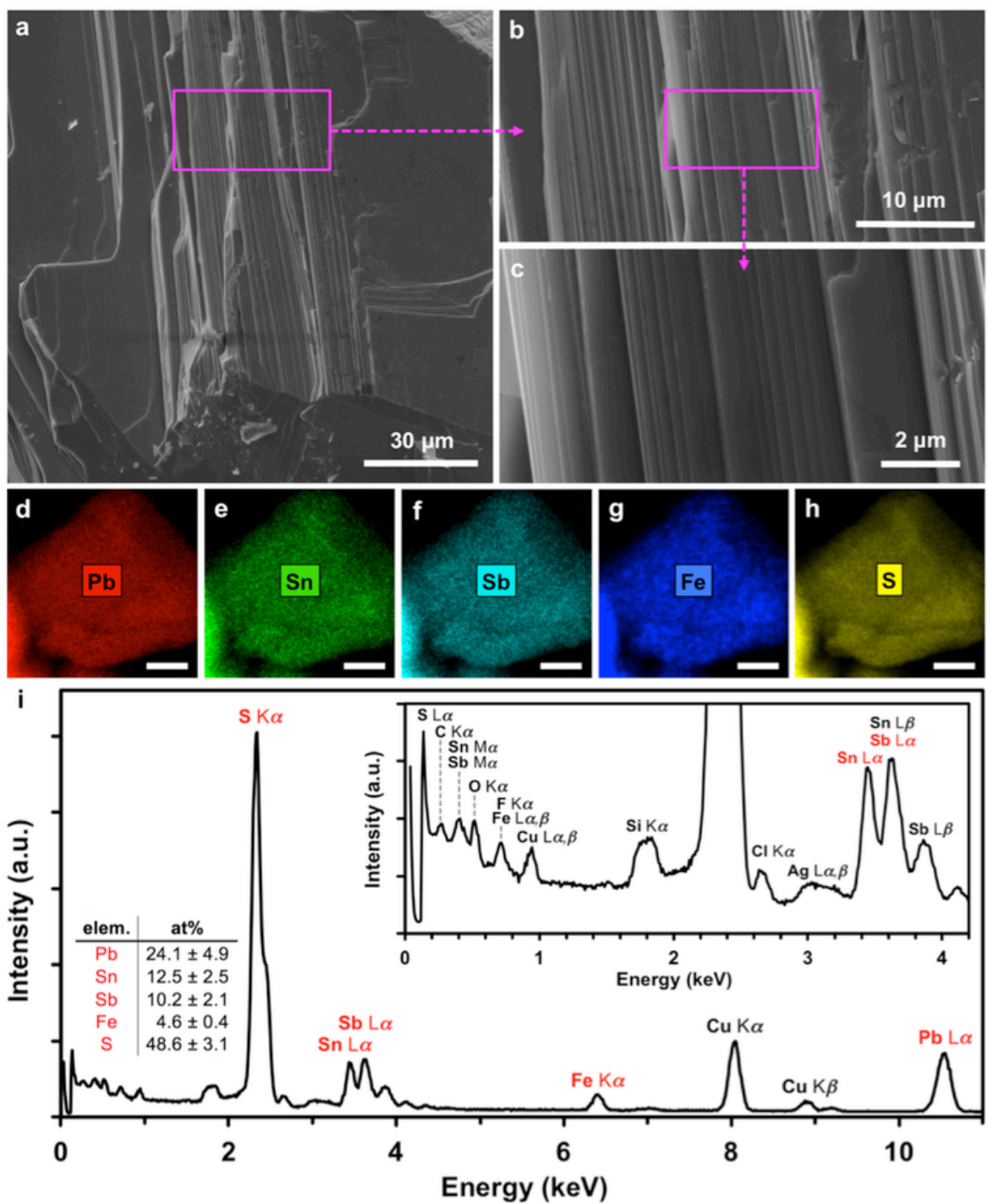

**Figure 1 | SEM and STEM-EDXS characterisation of franckeite. a,b,c**, SEM images of a franckeite crystal immobilised on a conductive carbon support. Zoom areas are highlighted by the magenta rectangles. **d−h**, STEM-EDXS maps for a franckeite crystal showing lead, tin, antimony, iron, and sulphur, respectively. The scale bars in **d−h** denote 30 nm. **i**, Averaged EDXS spectra with the inset showing the low-intensity peaks found at low energy. The peaks of the five major elements used for the quantification are marked in red. Secondary peaks and the peaks originating from the substrate and impurities, which were deconvoluted and not included in the quantification, are marked in black. Full quantification and further details are found in the Supplementary Information, section S1.

Figure 2 shows the high-angle annular dark-field (HAADF) STEM characterisation of franckeite. This mineral is a complex layered misfit compound, whose structure has not yet been fully determined.[9, 10, 11, 12] Our results confirm phase segregation into discrete Sn-rich pseudo-hexagonal (H) and Pb-rich pseudo-tetragonal (T) layers with a van der Waals gap between them. A cross-section through a thin exfoliated crystal was extracted to reveal a unit cell height of ~1.85 nm (Fig. 2b,e). The two layers, H and T, are incommensurate with differing lattice dimensions in the [010] direction, which results in a mismatch periodicity of ~4.3 nm visible as moiré fringes when viewed along the [001] direction in the TEM and also confirmed by the relative atomic displacements of the two layers in the cross-section (Supplementary Figure S2 and S3). Atomic-resolution HAADF STEM (Fig. 2e) and STEM EDXS elemental analysis of this structure (Fig. 2c,f) revealed significant compositional segregation at the atomic scale. Specifically, Pb in the inner two atomic planes of T layers is partially replaced by Sb resulting in an alternating Pb-Sb-Pb arrangement (Fig. 2d).

Interestingly, the cross-sectional STEM images also show stacking faults, consisting of irregular lateral displacements of the $SnS_2$ layers along the [010] direction, which is likely to cause the forbidden lattice spots, which appear in the electron diffraction data (Supplementary Figure S2 and S3). The thinnest region cross-sectioned for STEM analysis was ~6 nm or ~3 unit cells thick (Fig. 2e-f) with the upper (freshly cleaved) surface terminating at the $SnS_2$ layer and covered by ~0.5 nm thick layer of carbonaceous adsorbates. In contrast, the lower surface terminates part way through the Pb-rich layer. This suggests that the incommensurate stacking between the H and T layers results in predominantly uniaxial basal cleavage of franckeite. SnS and $SnS_2$ have been recently shown to be readily exfoliated[13, 14] and preferential elemental segregation to a single Pb-rich atomic plan within a $PbSnS_2$ lattice was also previously observed.[3] The observed phase segregation, which results in a van der Waals gap between the layers, proves the concept of a naturally occurring heterostructure material, which should be common among many mixed-metal sulphides and other crystals. This could be exploited in emerging synthesis of metal chalcogenides and tailoring of their properties.[15]

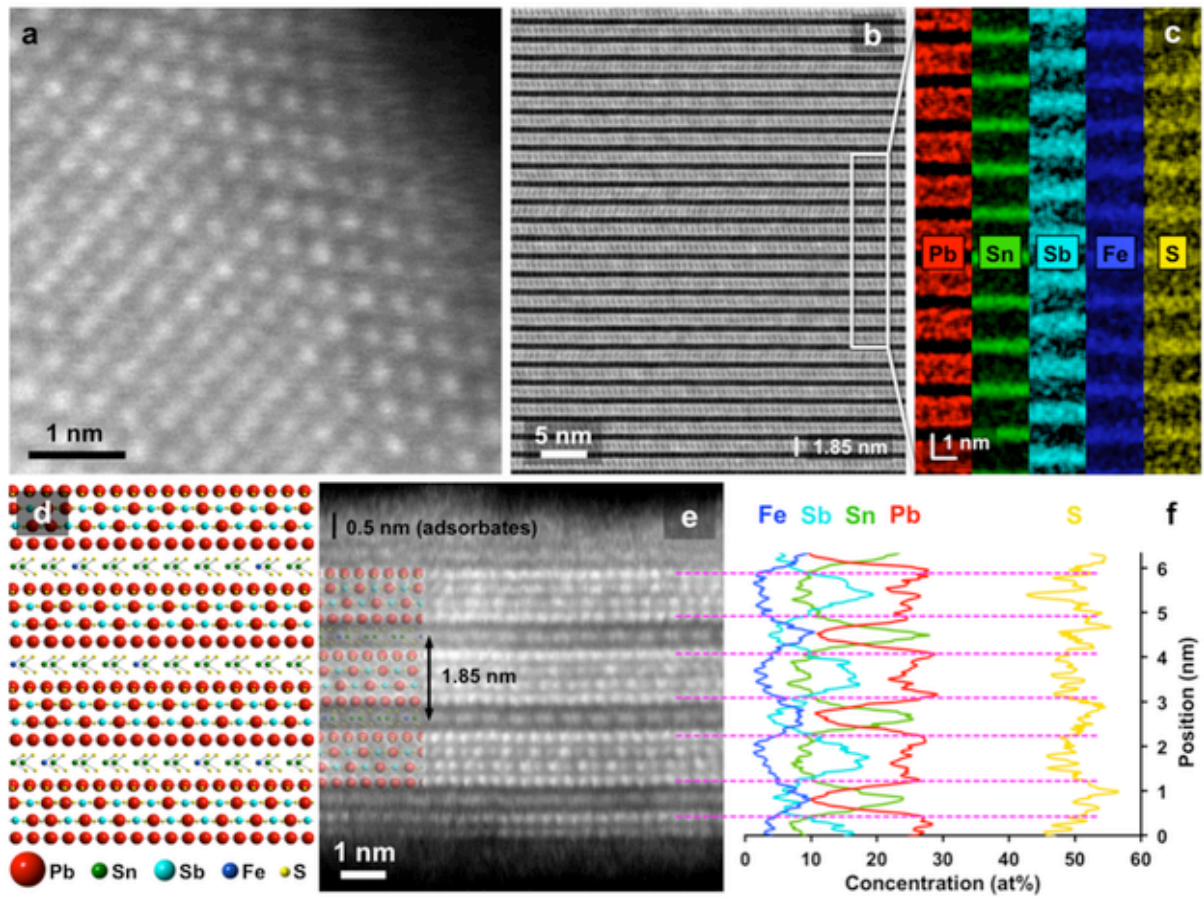

**Figure 2 | HAADF-STEM imaging of franckeite. a**, High-resolution image of the crystal lattice viewed along the [001] direction (plan view of the basal cleavage plane). **b**, Cross-sectional image of the layering viewed along the [100] direction. **c**, Corresponding EDXS elemental intensity maps. **d**, Proposed structure model of franckeite, with the relative size of atoms corresponding to their brightness in the HAADF image. **e**, Atomic-resolution image of the cross-section for a few-layer thick crystal. The inset schematically shows an overlay of franckeite structure. **f**, Corresponding EDXS concentration profiles showing compositional variation both between and within the individual layers. In contrast, no compositional variations were detected when the crystal is viewed along the [001] direction.

The surface sensitivity and a few-nanometre penetration depth of XPS were utilised to determine the surface composition of franckeite including the oxidation states and extent of surface impurities. Fig. 3a shows a wide-range XPS spectrum and quantification using the main elements (inset table). High-resolution spectra of Pb 4f, Sn 3d, Sb 3d, and S 2p binding energy regions are shown in Fig. 3b–e. Peak fitting was used to determine the oxidation states of individual elements. Lead and tin (Fig. 3b and 3c) are present as $Pb^{2+}$ and $Sn^{2+}$ (both ~80%), and $Pb^{4+}$ and $Sn^{4+}$ (both ~20 %). Antimony (Fig. 3d) is present as $Sb^{3+}$ (~60 %) and $Sb^{5+}$ (~40 %). Sulphur (Fig. 3e) is present exclusively as $S^{2-}$. Only a negligible amount of Fe (< 0.5 at%) was detected, and Sn concentration is significantly smaller than measured from EDXS. Composition stoichiometry analysis yielded an approximate surface composition chemical formula of $Pb_3SnSb_2S_6$. This suggests that the Sn-rich H layers, which also contain most of Fe (Fig. 2c,f), are underrepresented on the franckeite surface, possibly because the surface is unstable upon exfoliation and undergoes surface reconstruction or degradation.

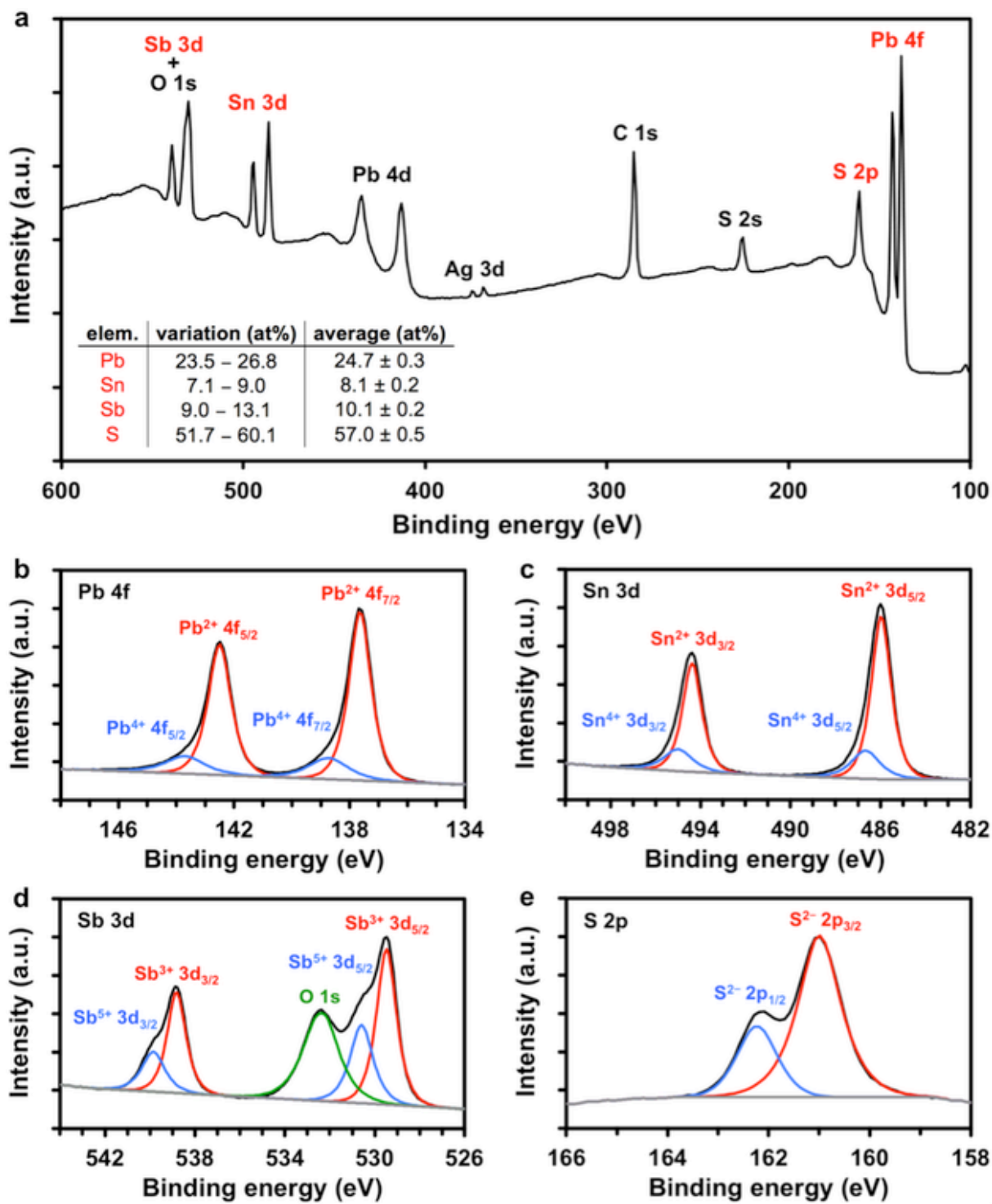

**Figure 3 | XPS analysis of franckeite surface. a**, Average XPS spectrum of franckeite obtained from multiple individual measurements, referenced to the adventitious carbon C 1s peak at 284.7 eV.[16] The peaks labelled in red were used for the quantification shown in the inset table. Other significant peaks, which were not used in quantification, are labelled in black. **b,c,d,e**, High-resolution spectra of Pb 4f, Sn 3d, Sb 3d (+ O 1s), and S 2p peaks, respectively. The intensities are normalised to the most intense peak within the respective spectral region. The stoichiometric analysis, full quantification, and comparison of freshly exfoliated and air-aged crystals can be found in Supplementary Information, section S3.

Importantly, we have succeeded in isolating monolayer, or more precisely single unit cell thick, franckeite crystals. Thin layers were mechanically exfoliated onto a $SiO_2$ substrate and characterised using optical microscopy, AFM, and Raman spectroscopy. Fig. 4a shows an optical micrograph of a bulk crystal with an adjacent few-layer region. The AFM height-contrast between the substrate, monolayer, few-layer, and bulk is clearly visible in Fig. 4b. The monolayers have typical thickness of ~2.4 nm, which is in good agreement with the single unit cell repeat period determined from cross-sectional TEM (~1.85 nm), considering the AFM instrumental offset and layer of carbonaceous adsorbates or trapped moisture, which are known to beset surfaces of 2D materials.[2, 17, 18] The length of few-layer crystals was typically tens of μm, but their width never increased above 1–2 μm, resulting in characteristic needle-like thin crystals. This preferential uniaxial basal cleavage is supported by the observed commensurate and incommensurate lattice stacking in the [100] and [010] directions, respectively. Raman spectra of bulk and monolayer franckeite obtained using 532 nm and 633 nm excitation wavelength lasers are shown in Fig. 4c and 4d, respectively. Despite the lack of literature on franckeite, we infer that its Raman spectrum is dominated by $Sb_2S_3$ (stibnite) and $SnS_2$ (berndtite) vibrations,[19, 20] with corresponding peaks near 260 $cm^{-1}$ and 320 $cm^{-1}$, respectively, which decrease in intensity for monolayer in comparison to bulk crystals. These two peaks are blue-shifted by ~3 $cm^{-1}$ and ~1 $cm^{-1}$ in monolayer for the red laser only. $Sb_2S_3$ and $SnS_2$ both have large optical band gaps of 1.72 and 2.10 eV,[4] respectively, and therefore exhibit comparatively weaker light absorption and stronger Raman scattering than PbS (0.37 eV),[4] which is a relatively weak Raman scatterer and does not contribute to the signal significantly.[21] A weak broad band, centred around 190 $cm^{-1}$ in bulk franckeite, which diminishes and blueshifts by about 10 – 15 $cm^{-1}$ in the monolayer for both excitation wavelengths, could originate from SnS (1.01 eV) vibrations.[4, 22] No significant photoluminescence was detected within the 532 – 800 nm range (using 532 nm excitation).

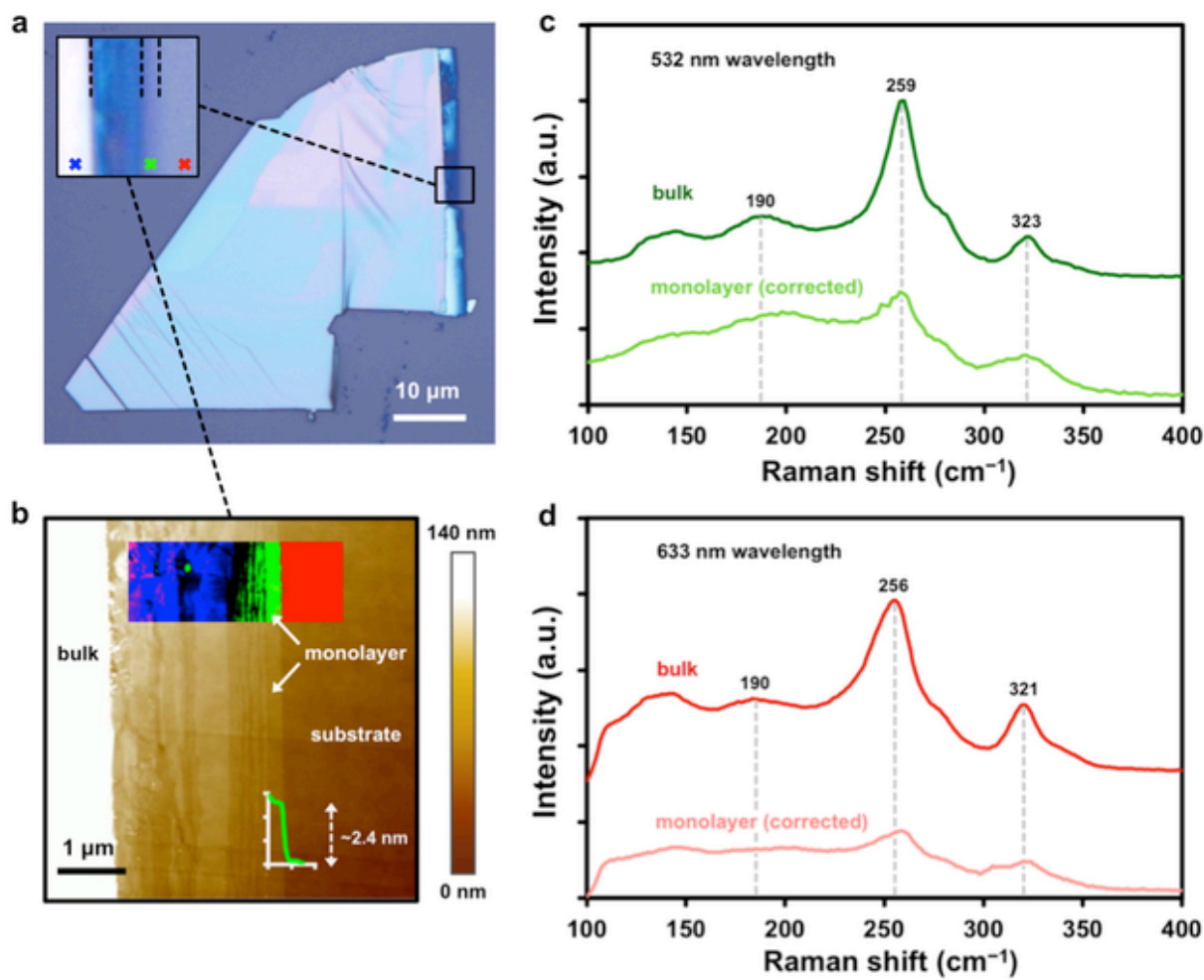

**Figure 4 | Identification of a single unit cell franckeite crystal using optical microscopy, AFM, and Raman spectroscopy. a**, Bright-field optical micrograph of a franckeite crystal exfoliated on $SiO_2$. The inset shows the thin crystal region, with the boundaries between bulk, few-layer, and monolayer crystals, and the $SiO_2$ substrate (left to right), outlined by dashed vertical lines. **b**, AFM image of the thin region. The discrete-colour inset highlights the height contrast between the regions of different thickness. **c,d**, Raman spectra of bulk and monolayer franckeite using a green (532 nm) and red (633 nm) laser, taken at locations indicated by the coloured crosses in the inset of **a**. The intensities are normalised to the most intense bulk peaks at 259 cm$^{-1}$ (**c**) and 256 cm$^{-1}$ (**d**). The background spectrum of the underlying $SiO_2$ substrate was subtracted from both monolayer spectra. Further characterisation using these techniques is detailed in Supplementary Information, section S4.

In order to explore the potential use of franckeite in energy storage and conversion applications, we have determined its electrochemical properties using a microdroplet cell measurement, representative results of which are summarised in Figure 5. This approach was previously applied to topography-dependent electrochemistry of graphene and $MoS_2$.[17, 23, 24] Since the cleavage of franckeite is prevalently limited to a single plane, we use the term 'basal' to describe their low-defect (001) surface, and term 'edge' to describe other surfaces with an increased density of edges, terraces, and defects. We have found that the averaged electric double-layer capacitance measured by cyclic voltammetry on the basal surface is 27.4 ± 2.2 μF $cm^{-2}$ and further increases to 153 ± 62.8 μF $cm^{-2}$ on the edge surface (Fig. 5a). In comparison, typical capacitance values measured using the same method were ca. 15 – 30 μF $cm^{-2}$ on an unpolished platinum surface, 1 – 2 μF $cm^{-2}$ on basal plane graphene, and 2 – 4 μF $cm^{-2}$ on basal plane $MoS_2$. Such high capacitance indicates a degenerate semiconducting nature of franckeite and an additional contribution from redox activity, which can be observed when the potential window is extended (Supplementary Figure S10). We propose that this is directly related to its electronic structure as it has been previously proposed for $MoS_2$ that steric accessibility of metallic orbitals at the crystal edges increases its electrochemical performance.[25] We cannot completely rule out that changes in the active area of the edge surface, or even solution-induced delamination of the crystals, contribute to the capacitance increase. However, if this were the case, it would in fact aid liquid-exfoliation and ion intercalation of franckeite.

Heterogeneous electron transfer rate measurement using two common redox mediators, $Ru[(NH_3)_6]^{3+/2+}$ and $[Fe(CN)_6]^{3-/4-}$, is shown in Fig. 5c, yielding average values of (0.62 ± 0.34) × $10^{-3}$ cm $s^{-1}$ for $Ru[(NH_3)_6]^{3+/2+}$ and (0.90 ± 0.17) × $10^{-3}$ cm $s^{-1}$ for $[Fe(CN)_6]^{3-/4-}$. These values exceed those determined on graphite and $MoS_2$ using the same experimental method.[17] Hydrogen evolution, an important technological reaction, was difficult to quantify in a diffusion-limited regime within the microdroplet electrochemical cell. Nevertheless, some hydrogen evolution activity was observed on basal surface in 1 M HCl solution as compared to the blank electrolyte solution (Fig. 5d).

We also determined electronic transport characteristics of bilayer and 5-layer franckeite as shown in Fig. 5d-f. Both thicknesses display electric field-effect behaviour, with the bilayer device showing stronger conductance modulation, cf. top insets on Fig. 5d and 5e. From the gate voltage dependence we conclude that the material is a p-doped semiconductor. Current-bias voltage characteristics were nonlinear suggesting the formation of Schottky barriers at the interface between the metal and franckeite, also supported by a strong temperature dependence of current-voltage characteristics (Fig. 5f). The activation energies extracted from Arrhenius plots were 220 meV and 80 meV for bilayer and 5-layer devices, respectively.

In summary, we have successfully exfoliated franckeite crystals to a single unit cell thickness, which is facilitated through phase segregation into discrete layers at the nanometre scale. The capacitance and electron transfer rate are considerably higher than those of more common 2D materials. In particular, the capacitance of franckeite exceeds that of graphene and $MoS_2$ by one order of magnitude. The pseudocapacitive character of franckeite and possibility to exfoliate it to thin, needle-like monolayer and few-layer crystals has a great potential for liquid-phase exfoliation and could find use in emerging energy-storage technologies such as supercapacitors. The electronic transport measurements show that franckeite is a p-doped semiconductor. Overall, these findings open the way to a new class of complex van der Waals crystals with prevalent basal cleavage as an alternative to artificially stacked 2D heterostructures.[26]

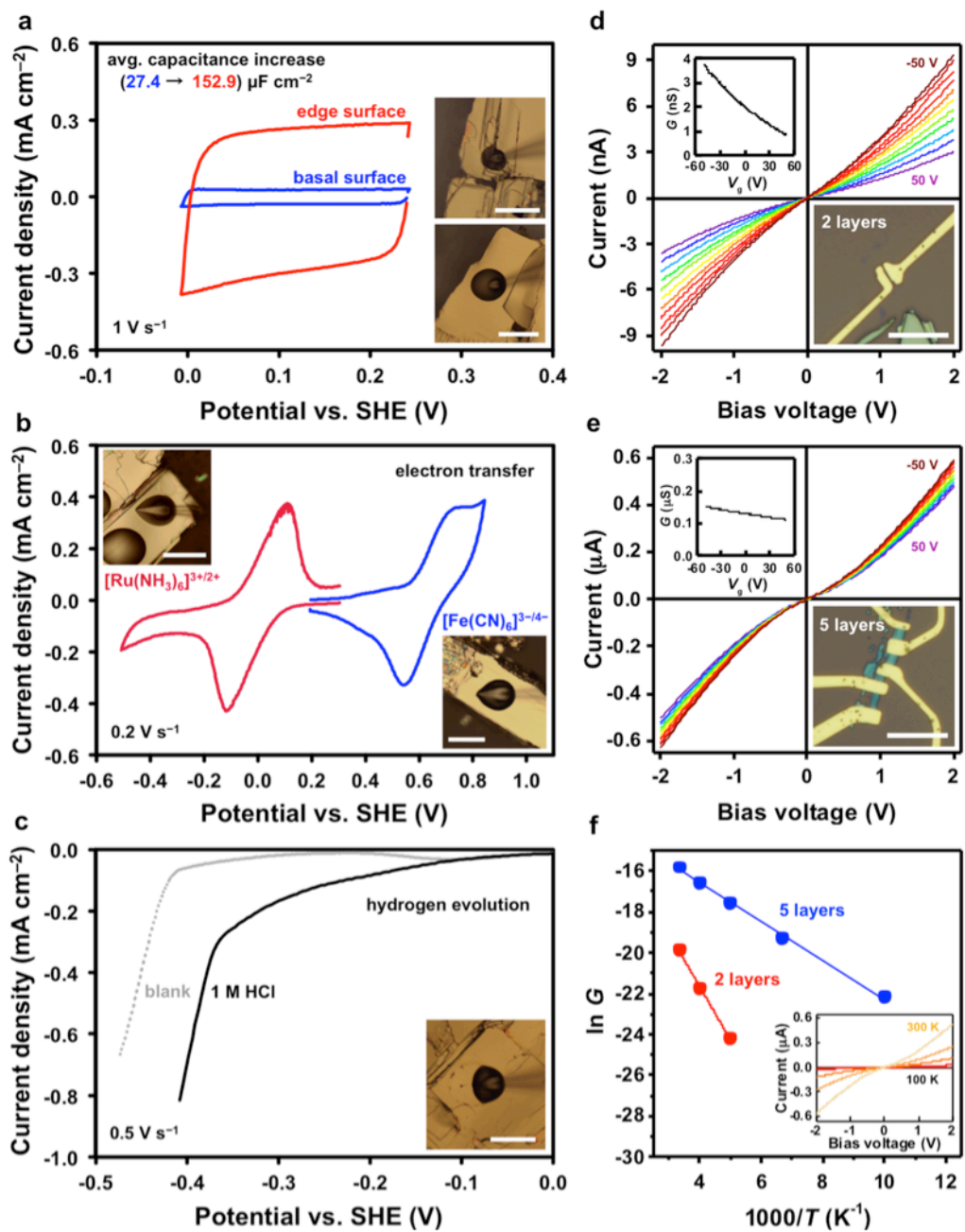

**Figure 5 | Electrochemical and electronic transport measurements on franckeite. a**, Capacitance measurement in 6 M LiCl on basal and edge surface. **b**, Electron transfer measurement on basal surface using $Ru[(NH_3)_6]^{3+/2+}$ and $[Fe(CN)_6]^{3-/4-}$ in 6 M LiCl. **c**, Hydrogen evolution measurement using 1 M HCl in 6 M LiCl. The optical micrographs in the inset show the microdroplet cells, scale bars are 50 μm. Further electrochemical data are found in Supplementary Information, section S5. **d**, Gate voltage ($V_g$) dependence of the current-voltage for bilayer franckeite between −50 to 50 V (brown to violet), in 10 V increments at 300 K. Top-left inset: conductance ($G$) dependence on $V_g$ (300 K). **e**, Same as **d** for a ~5-layer franckeite. Bottom-right insets: optical micrographs of the devices, scale bars correspond to 10 μm. **f**, Temperature dependence of zero-bias conductance for bilayer (red) and 5-layer (blue), solid lines are Arrhenius fits. The inset shows current-voltage curves for the 5-layer device between 300 and 100 K (yellow to black) in 50 K increments.

**Methods**

**Sample preparation**

The samples were prepared by the 'scotch-tape' mechanical exfoliation of natural franckeite crystals originating from Poopo, Oruro Department, Bolivia (Manchester Nanomaterials Ltd, UK) onto three different types of substrate: carbon adhesive discs for SEM and XPS analysis, lacey carbon-coated copper grids for STEM, TEM and EDXS analysis (both Agar Scientific, UK), and oxidised silicon wafers (IBD Technologies, UK) for electrochemical, electronic transport measurements, Raman spectroscopy, and focused ion beam (FIB) milling. The two-terminal devices for the transport measurement were prepared using e-beam lithography followed by e-beam evaporation of Cr/Au contacts (insets of Fig. 5d-e). Exfoliated crystals were stable in air for at least 6 months, based on no observable changes in optical microscopy, Raman spectroscopy, and AFM.

**SEM and XPS characterisation**

SEM images were collected using Philips XL30 ESEM-FEG scanning electron microscope (FEI Company, USA) operating at 15 kV accelerating voltage. XPS spectra were obtained using a K-Alpha monochromated XPS spectrometer (Thermo Fisher Scientific Inc) and analysed using CasaXPS software v.2.3 (Casa Software Ltd).

**TEM/STEM characterisation**

Electron diffraction patterns were collected using a Philips CM20 TEM operated at 200 kV accelerating voltage. STEM images and EDXS elemental maps were obtained using a Titan G2 STEM (FEI Company, USA) operated at 200 kV, equipped with a Super-X EDX detector and GIF quantum energy filter. HAADF images were acquired with a convergence angle of 21 mrad, an inner angle of 54 mrad and a probe current of ~75 pA, EDXS data was quantified using Esprit software version 1.9 (Bruker, USA). The crystals were aligned using Kikuchi bands in the $SiO_2$ substrate. Cross-sectional sample preparation was performed using a dual FIB Nova NanoLab instrument (FEI Company, USA) fitted with an Omniprobe™ nano-manipulator (Oxford Instruments, UK). More details of milling procedures can be found in Supplementary information, section S2.

**Optical microscopy, AFM, and Raman spectroscopy**

A Nikon Eclipse LV100ND optical microscope and a DS-Fi2 U3 CCD camera (Nikon Metrology, UK Ltd.) were used to image the flakes using both bright-field and dark-field illumination modes. AFM thickness measurements were performed with the Bruker Dimension 3100V instrument in a tapping mode with a tip resonance frequency of ~350 kHz. Raman spectroscopy was measured using an inVia spectrometer with 532 and 633 nm excitation wavelength lasers of variable power density, 1800 grooves/mm grating and a 100× objective (Renishaw plc, UK), resulting in a spot size of ca. 0.8 $\mu m^2$.

**Electrochemical measurements**

The franckeite crystals were electrically contacted to a copper wire using a silver conductive paint (RS Components Ltd, UK). Aqueous droplets of either pure 6 M LiCl electrolyte, 3 mM redox mediator ($Ru(NH_3)_6Cl_3$ or $K_3Fe(CN)_6$) in 6 M LiCl, or 1 M hydrochloric acid in 6 M LiCl, deposited onto the flake surface using a pressure-controlled glass micropipette, were used as microscopic electrochemical cells. Ultrapure deionised water (18.2 MΩ cm, Milli-Q Direct 8, Merck Millipore, USA) was used for preparation of these solutions. Electrochemical measurements were controlled by a PGSTAT302N potentiostat (Metrohm Autolab B.V., The Netherlands) and were carried out in a three-electrode configuration, employing the crystal surface as a working electrode, and platinum wire and silver chloride wire as a counter and a reference electrode, respectively. All measurements were performed at room temperature (25 − 28°C) and the potential is referenced to the standard hydrogen electrode (SHE). All chemicals were of 98 % or higher purity and were purchased from Sigma-Aldrich, UK. Copper, silver and platinum wires (>99.9 %) were purchased from Advent Research Materials, UK. Further details can be found in the Supplementary Information, section S5, and in our previous publications.[17, 23, 24]

**Electrochemical analysis**

The heterogeneous electron transfer rate constant, $k^0$, was calculated from the following equation.[27]

$$k^0 = 2.18\left(\frac{\alpha nFDv}{RT}\right)^{0.5} e^{-\frac{\alpha^2 nF}{RT}\Delta E_p} \qquad (1)$$

where $\alpha$ (= 0.5 due to the reaction symmetry) is the transfer coefficient, $n = 1$ is the number of electrons exchanged in the reaction, $F$ is the Faraday constant, $D$ is the diffusion coefficient of the redox mediator, $v$ is the scan rate, $R$ is the universal gas constant, $T$ is the thermodynamic temperature, and $\Delta E_p$ (larger than 220 mV) is the peak-to-peak separation of a redox mediator reduction/oxidation reaction. The Nicholson method, based on the following equation, was used for $\Delta E_p$ smaller than 220 mV.[28, 29]

$$\psi = \frac{(-0.629 + 0.002 n\Delta E_p)}{(1 - 0.017 n\Delta E_p)} = k^0 \sqrt{\frac{RT}{\pi nFD}} v^{-0.5} \qquad (2)$$

$\Delta E_p$ was measured for each microdroplet for a range of scan rates (0.1 – 1 V s$^{-1}$) and mean $k^0$ value determined using the equations (1) and (2).

The electric double-layer capacitance, $C_{EDL}$, was determined from a cyclic voltammetry using the following equation.[30]

$$C_{EDL} = \frac{1}{2v(E_{max} - E_{min})} \oint_E I(E) dE \qquad (3)$$

where $E$ is the applied potential and $E_{max}(E_{min})$ are the maximum(minimum) potentials, which limit the voltammetric scan. The mean $C_{EDL}$ was determined from several measurements at different scan rates within the range of 0.3 – 3 V s$^{-1}$.

# References

1. Novoselov K. S., *et al.* Two-dimensional atomic crystals. *Proc. Natl. Acad. Sci. U. S. A.* **102**, 10451-10453 (2005).
2. Butler S. Z., *et al.* Progress, Challenges, and Opportunities in Two-Dimensional Materials Beyond Graphene. *ACS Nano* **7**, 2898-2926 (2013).
3. He J., Girard S. N., Zheng J. C., Zhao L., Kanatzidis M. G.& Dravid V. P. Strong phonon scattering by layer structured $PbSnS_2$ in PbTe based thermoelectric materials. *Adv. Mater.* **24**, 4440-4444 (2012).
4. Xu Y.& Schoonen M. A. A. The absolute energy positions of conduction and valence bands of selected semiconducting minerals. *Am. Mineral.* **85**, 543-556 (2000).
5. Boldish S. I.& White W. B. Optical band gaps of selected ternary sulfide minerals. *Am. Mineral.* **83**, 865-871 (1998).
6. Unuchak D. M., *et al.* Structure and optical properties of PbS-SnS mixed crystal thin films. *Phys. Status Solidi C* **6**, 1191-1194 (2009).
7. Soriano R. B., Malliakas C. D., Wu J.& Kanatzidis M. G. Cubic Form of $Pb_{2-x}Sn_xS_2$ Stabilized through Size Reduction to the Nanoscale. *J. Am. Chem. Soc.* **134**, 3228-3233 (2012).
8. Nair P. K., Nair M. T. S., Fernandez A.& Ocampo M. Prospects of chemically deposited metal chalcogenide thin films for solar control applications. *J. Phys. D: Appl. Phys.* **22**, 829-836 (1989).
9. Moh G. H. Mutual $Pb^{2+}/Sn^{2+}$ substitution in sulfosalts. *Mineral. Petrol.* **36**, 191-204 (1987).
10. Williams T. B.& Hyde B. G. Electron microscopy of cylindrite and franckeite. *Phys. Chem. Miner.* **15**, 521-544 (1988).
11. Wang S.& Kuo K. H. Crystal lattices and crystal chemistry of cylindrite and franckeite. *Acta Crystallogr., Sect. A: Found. Crystallogr.* **47**, 381-392 (1991).
12. Wang S.& Buseck P. R. High-angle annular dark-field microscopy of franckeite. *Am. Mineral.* **80**, 1174-1178 (1995).
13. Brent J. R., *et al.* Tin(II) Sulfide (SnS) Nanosheets by Liquid-Phase Exfoliation of Herzenbergite: IV–VI Main Group Two-Dimensional Atomic Crystals. *J. Am. Chem. Soc.* **137**, 12689-12696 (2015).
14. Burton L. A., *et al.* Electronic and optical properties of single crystal $SnS_2$: an earth-abundant disulfide photocatalyst. *J. Mater. Chem. A* **4**, 1312-1318 (2016).
15. Feng W., *et al.* Synthesis, properties and applications of 2D non-graphene materials. *Nanotechnology* **26**, 292001 (2015).
16. Barr T. L.& Seal S. Nature of the use of adventitious carbon as a binding energy standard. *J. Vac. Sci. Technol., A* **13**, 1239-1246 (1995).
17. Velický M., *et al.* Electron transfer kinetics on natural crystals of $MoS_2$ and graphite. *Phys. Chem. Chem. Phys.* **17**, 17844-17853 (2015).
18. Li Z., *et al.* Water Protects Graphitic Surface from Airborne Hydrocarbon Contamination. *ACS Nano* **10**, 349-359 (2016).
19. Mead D. G.& Irwin J. C. Raman spectra of $SnS_2$ and $SnSe_2$. *Solid State Commun.* **20**, 885-887 (1976).
20. Makreski P., Petruševski G., Ugarković S.& Jovanovski G. Laser-induced transformation of stibnite ($Sb_2S_3$) and other structurally related salts. *Vib. Spectrosc.* **68**, 177-182 (2013).
21. Smith G. D., Firth S., Clark R. J. H.& Cardona M. First- and second-order Raman spectra of galena (PbS). *J. Appl. Phys.* **92**, 4375-4380 (2002).

22. Chandrasekhar H. R., Humphreys R. G., Zwick U. & Cardona M. Infrared and Raman spectra of the IV-VI compounds SnS and SnSe. *Physical Review B* **15**, 2177-2183 (1977).
23. Velický M., *et al.* Electron Transfer Kinetics on Mono- and Multilayer Graphene. *ACS Nano* **8**, 10089-10100 (2014).
24. Velický M., *et al.* Photoelectrochemistry of Pristine Mono- and Few-Layer $MoS_2$. *Nano Lett.* **16**, 2023-2032 (2016).
25. Ahmed S. M. & Gerischer H. Influence of crystal surface orientation on redox reactions at semiconducting $MoS_2$. *Electrochim. Acta* **24**, 705-711 (1979).
26. Withers F., *et al.* Light-emitting diodes by band-structure engineering in van der Waals heterostructures. *Nat. Mater.* **14**, 301-306 (2015).
27. Klingler R. J. & Kochi J. K. Electron-transfer kinetics from cyclic voltammetry. Quantitative description of electrochemical reversibility. *J. Phys. Chem.* **85**, 1731-1741 (1981).
28. Nicholson R. S. Theory and application of cyclic voltammetry for measurement of electrode reaction kinetics. *Anal. Chem.* **37**, 1351-1355 (1965).
29. Lavagnini I., Antiochia R. & Magno F. An extended method for the practical evaluation of the standard rate constant from cyclic voltammetric data. *Electroanalysis* **16**, 505-506 (2004).
30. Xiong G., Meng C., Reifenberger R. G., Irazoqui P. P. & Fisher T. S. A review of graphene-based electrochemical microsupercapacitors. *Electroanalysis* **26**, 30-51 (2014).

**Acknowledgments**

This work was supported by EPSRC (grants EP/I005145/1, EP/K016954/1, EP/K016946/1, EP/M010619/1, EP/L01548X/1, EP/G03737X/1, and EP/N007131/1), ERC, Graphene Flagship (contract No. NECTICT-604391), the Royal Society, the Defense Threat Reduction Agency (grant HDTRA1-12-1-0013), US Army Research Office, US Navy Research Office, and US Airforce Research Office. The Titan was funded by H.M government UK. Authors also thank NEXUS at nanoLAB (Newcastle University) for the XPS measurement.

**Author contributions**

M.V. prepared the samples, carried out the electrochemical and optical characterisation, analysed the experimental data and wrote the manuscript, P.S.T. carried out SEM and Raman spectroscopy characterisation, S.J.H, A.P.R, and A.M.R carried out the FIB, TEM, STEM and EDXS measurements, A.K, C.R.W, and T.G contributed to the sample preparation and AFM measurements. A.M. performed the electronic transport measurements. S.J.H., K.S.N., and R.A.W.D. equally contributed to the discussion and interpretation of the results.

**Additional Information**

Supplementary information is available in the online version of the paper. Reprints and permissions information is available online at www.nature.com/reprints. Correspondence and requests for materials should be addressed to M.V or R.A.W.D.

**Competing financial interests**

The authors declare no competing financial interests.

# Supplementary Information:

# Phase segregation facilitates exfoliation of franckeite crystals to a single unit cell thickness

Matěj Velický,*[1] Peter S. Toth,[1] Alexander M. Rakowski,[2] Aidan P. Rooney,[2] Aleksey Kozikov,[3] Artem Mishchenko,[3] Colin R. Woods,[3] Thanasis Georgiou,[4] Sarah J. Haigh,[2] Kostya S. Novoselov,[3] and Robert A.W. Dryfe*[1]

[1] School of Chemistry, University of Manchester, Oxford Rd, Manchester, M13 9PL, UK

[2] School of Materials, University of Manchester, Oxford Rd, Manchester, M13 9PL, UK

[3] School of Physics and Astronomy, University of Manchester, Oxford Rd, Manchester, M13 9PL, UK.

[4] Manchester Nanomaterials Ltd, 83 Ducie Street, Manchester, M1 2JQ, UK

* To whom correspondence should be addressed. Tel: +44 (0)161-306-4522; Fax: +44 (0)161-275-4598, e-mail: matej.velicky@manchester.ac.uk or robert.dryfe@manchester.ac.uk

### Supplementary information content:

**S1. Scanning electron microscopy and energy-dispersive X-ray spectroscopy** (Figure S1, Table S1)

**S2. Scanning transmission electron microscopy** (Figure S2 –S3)

**S3. X-ray photoelectron spectroscopy** (Figures S4–S5, Tables S2–S3)

**S4. Raman spectroscopy, optical microscopy, and atomic force microscopy** (Figures S6–S8)

**S5. Electrochemical measurements** (Figures S9–S10)

**S1. Scanning electron microscopy and energy-dispersive X-ray spectroscopy**

Additional scanning electron microscopy (SEM) images of franckeite crystals at four different magnifications are shown in Fig. S1a−d. The low-magnification image (Fig. S1a) shows the 'crumbly' character of these crystals at millimetre scale. The high-magnification images (Fig. S1b−d) reveal a layered character at the micrometre scale, which facilitates mechanical exfoliation to thin needle-like, few single unit cell thick crystals, thus yielding a large surface area material for applications in energy storage, conversion, and catalysis.

Transmission electron microscope (TEM) energy-dispersive X-ray spectroscopy (EDXS) mapping of the elemental distribution within a franckeite crystal is shown in Fig. S1e−l. The EDXS maps reveal that Pb (Fig. S1e), Sn (Fig. S1f), Sb (Fig. S1g), Fe (Fig. S1h) and S (Fig. S1i), all display homogeneous spatial distribution, and closely follow the topography of the crystal observed in the bright-field (BF)–TEM image (Fig. S1l). This implies that all the metals and sulphur are evenly distributed throughout the crystal. On the other hand, this distribution-topography correlation is completely absent for carbon (Fig. S1j). This confirms that carbon is purely a surface contaminant. Oxygen shows a weaker distribution-topography correlation than the metals and sulphur (Fig. S1k), which suggests that oxygen may be present both as metal oxide and as surface organic contamination. This is also evidenced by the X-ray photoelectron spectroscopy (XPS) – see section S3.

Table S1 shows the average elemental composition determined from the entire EDXS spectrum image, including the main impurity elements. Carbon concentration (~8 at%) is much lower than that determined from XPS (~50 at%, Fig. S4), yet again confirming that carbon is present purely as a surface contaminant. On the other hand, the relative silver concentration is similar for both techniques, suggesting that it is a bulk contaminant.

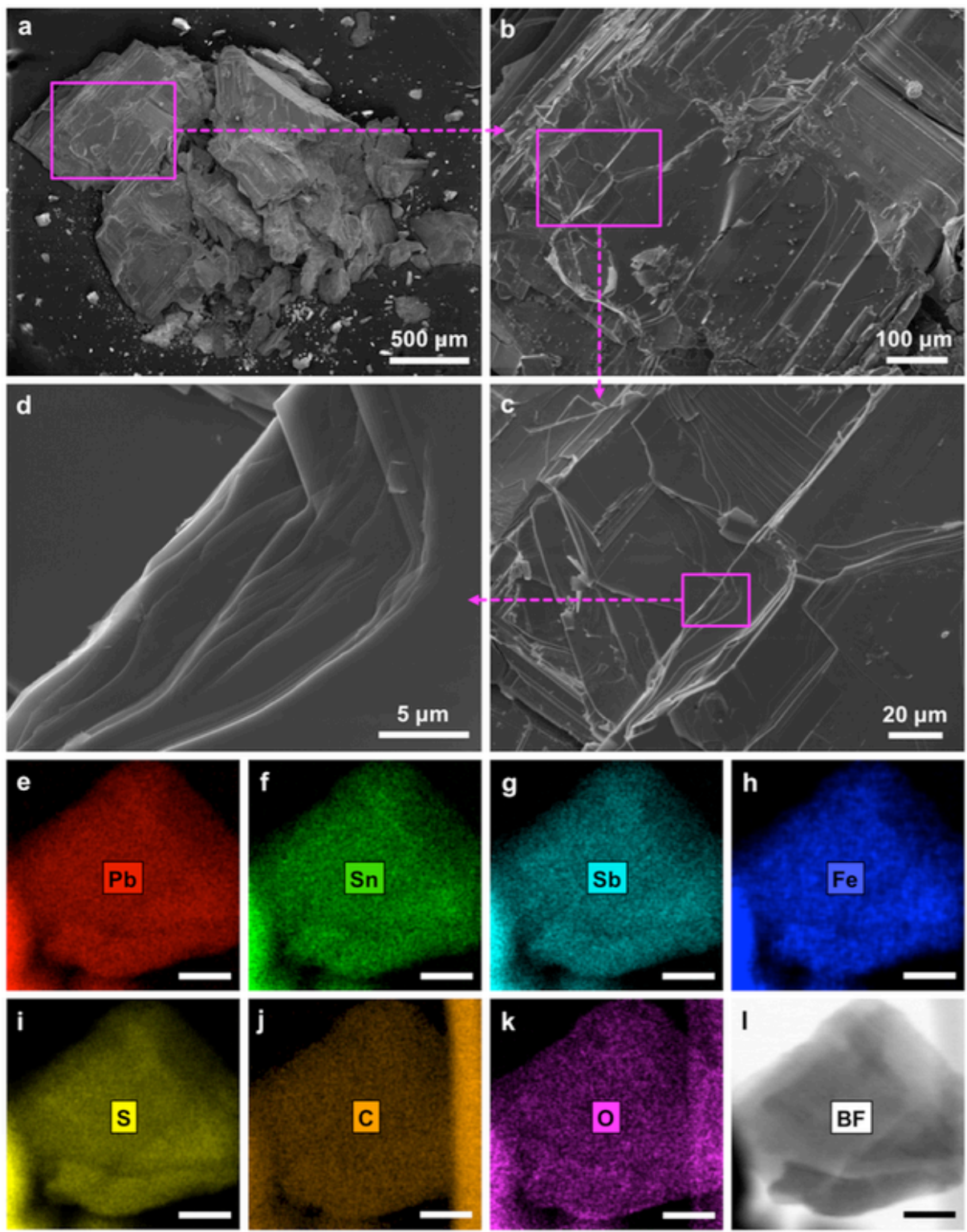

**Figure S1 | SEM images and TEM-EDXS mapping of franckeite. a**,**b**,**c**,**d**, SEM images of a crystal immobilised on a conductive carbon support. Magnified regions in **a**−**c** are highlighted by the magenta rectangles. **e**−**k**, EDXS maps of lead, tin, antimony, iron, sulphur, carbon, and oxygen, respectively, within another crystal. **l**, BF-TEM image of the same crystal. The scale bars in **e**–**l** denote 30 nm. Electron beam accelerating voltages of 15 kV and 200 kV were used in the SEM and TEM-EDXS measurement, respectively.

**Table S1 | EDXS quantification of franckeite crystals.**

| element | quantity / at% |
|---:|:---:|
| Pb | 21.1 ± 4.2 |
| Sn | 10.9 ± 2.2 |
| Sb | 8.9 ± 1.8 |
| Fe | 4.0 ± 0.3 |
| S | 42.5 ± 2.7 |
| O* | 3.7 ± 0.5 |
| C* | 7.7 ± 0.9 |
| Ag* | 1.1 ± 0.3 |

*bulk impurities and surface contamination

## S2. Scanning transmission electron microscopy

Figure S2 shows the indexed electron diffraction pattern and high-resolution TEM/scanning electron microscopy (STEM) images along the [001] direction, which reveal distinct moiré fringes originating from the H (Sn-rich) and T (Pb-rich) phase lattice mismatch, discussed in the main text. We also note that it was challenging to isolate a good franckeite single crystal for electron diffraction,[1] and the pattern in Fig. S2a contains contributions from more than one crystal.

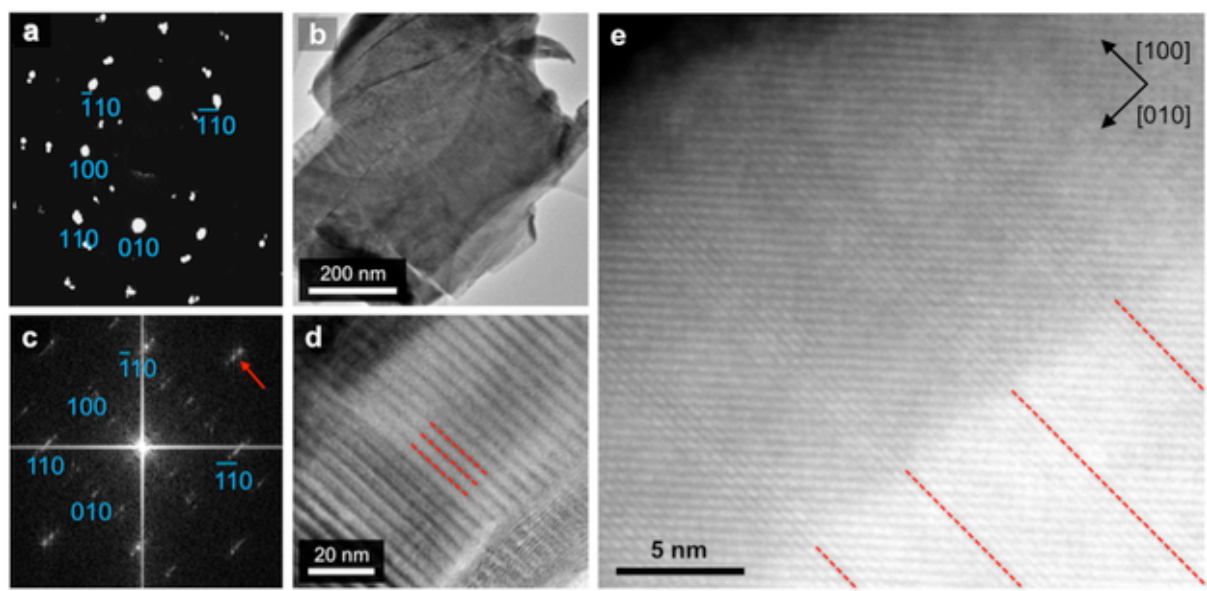

**Figure S2 | TEM imaging of the franckeite structure viewed along the [001] direction** (perpendicular to the basal plane). **a,** Indexed electron diffraction pattern. **b,d,e,** TEM image of a crystal viewed along the [001] direction. **c,** corresponding Fourier transform from the high resolution STEM image in **e**. These TEM images clearly show the ~4.4 nm moiré fringes, highlighted by the red dashed lines in (**d**) and (**e**). These fringes are also present in the Fourier transform image, causing the additional spots indicated by the red arrow in **c**.

The sample for cross-sectional characterisation was prepared using focused ion beam (FIB) milling as shown schematically in Fig. S3a. An ex situ sputter-coating of amorphous carbon (7 nm film) was followed by an Au-Pd layer (2 nm film) (Q150T, Quorum Technologies Ltd, Lewes, UK). A dual-beam FIB (Nova NanoLab, FEI) with FEG-SEM and gallium FIB columns were then used for a cross-section preparation with an in-situ lift out approach. The region of interest was identified using an SEM and a Pt protective strip (500 nm high) was formed using electron beam-induced deposition of trimethyl (methylcyclopentadienyl) platinum(IV), followed by a FIB Pt deposition (1.5 μm high).

Trenches were milled (30 kV, 10-1 nA) until the lamella thickness was 1 – 1.5 μm. Then, an in-situ lift out was performed using a micromanipulator (Omniprobe™), which is attached to the protective Pt layer of the cross-section, with further Pt deposition. The lamella was then transferred to a three-post FIB grid and thinned (30 kV at 0.5 – 0.1 nA, 5 kV at 50 pA and finally 2 kV at 90 pA) until electron transparent (20-50 nm thickness).[2] The moiré pattern observed along the [001] direction (Fig. S2d,e) is rationalised by high-angle annular dark-field (HAADF) STEM imaging in Fig. S3b, which shows an incommensurate stacking of the Sn-rich and Pb-rich layers in respect to one another. The crystal unit cells of these two layers along the [010] direction have different dimensions and only come into phase every ~4.4 nm (giving rise to the moiré fringes shown in Fig S2d,e). Fig. S3c shows the relative displacement of Sn-rich layers and Fig. S3d shows substitution of Sn-rich layers with a continuous PbS layer under the electron beam.

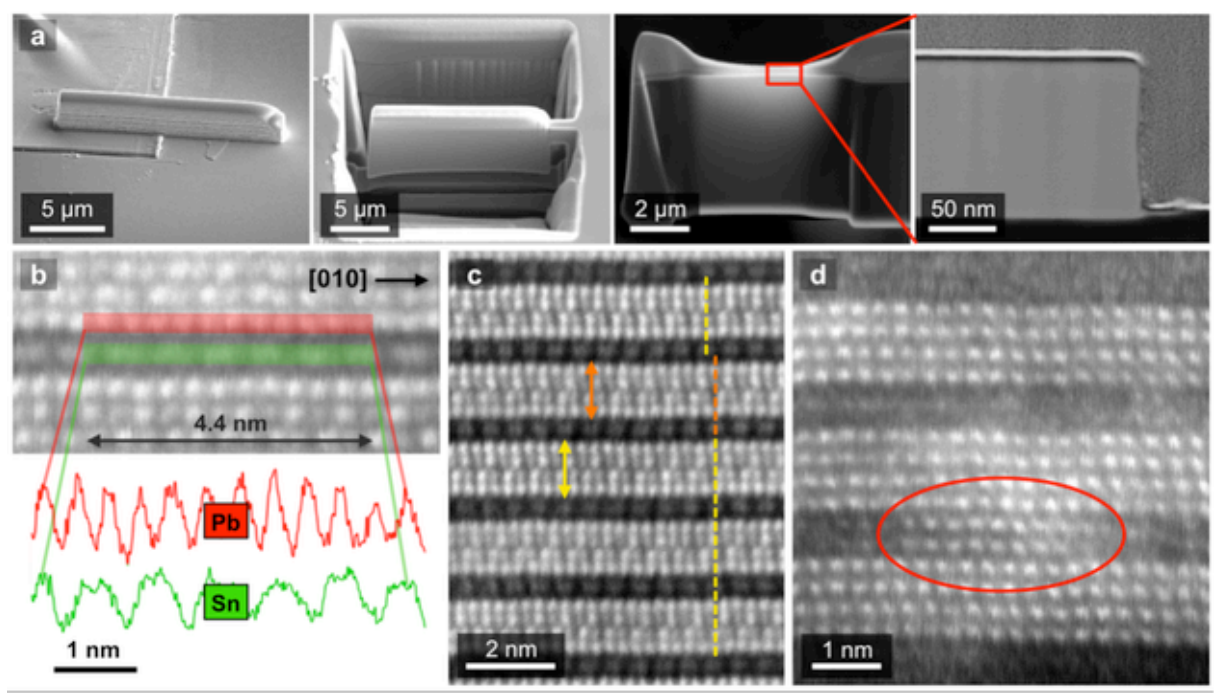

**Figure S3│ Cross-sectional HAADF-STEM imaging of franckeite along the [010] direction** (parallel to the basal plane). **a**, SEM and STEM images illustrating the process of FIB cross-sectional sample preparation from a crystal mechanically exfoliated on a $SiO_2$ substrate. **b**, Intensity line profiles reveal the incommensurate stacking between the $SnS_2$ phase (green) and the adjacent Pb(Sb)S phase (red). **c**, Variations in stacking for consecutive $SnS_2$ layers. The yellow arrow separates the $SnS_2$ layers in which the Sn atoms are positioned on top of one another, whereas the orange arrow indicates $SnS_2$ layers that have a relative displacement along the [010] direction. **d**, Electron beam induced restructuring of the $SnS_2$ layer, which forms a continuous PbS link between Pb-rich layers.

## S3. X-ray photoelectron spectroscopy

The composition stoichiometry of franckeite surface, derived from XPS data, was determined using equations (S1–S4). This analysis was based on the balance between the atomic fractions and oxidation states of lead, tin, antimony, sulphur, and oxygen. Equation (S1) expresses the material balance of different elements within franckeite surface.

$$\sum_i x_i = 1 \qquad (S1)$$

$x_i$ is the atomic fraction of an element $i$ (i.e. Pb or Sn) determined from XPS quantification. Equation (S2) expresses the material balance of different species of the same element.

$$\sum_j f_j = 1 \qquad (S2)$$

$f_j$ is the fraction of a species $j$ (i.e. $Pb^{2+}$ or $Pb^{4+}$) determined from high-resolution XPS spectra. Equation (S3) expresses the charge neutrality condition for the chemical formula.

$$\sum_i \sum_j x_i f_j z_j = 0 \qquad (S3)$$

$z_j$ is the charge number of a species $j$ (i.e. −2 for $S^{2-}$). The atomic fractions, species fractions, and species charge numbers are listed in Table S2.

**Table S2 | Composition stoichiometry parameters determined from the XPS data.**

| species | $x_i$ | $f_j$ | $z_j$ |
|---|---|---|---|
| $Pb^{2+}$ | 0.238 | 0.815 | 2 |
| $Pb^{4+}$ |  | 0.185 | 4 |
| $Sn^{2+}$ | 0.078 | 0.785 | 2 |
| $Sn^{4+}$ |  | 0.215 | 4 |
| $Sb^{3+}$ | 0.097 | 0.637 | 3 |
| $Sb^{5+}$ |  | 0.363 | 5 |
| $O^{2-}$ | 0.038 | 0.236* | −2 |
| $O_{(carb.)}$ |  | 0.764 | N/A |
| $S^{2-}$ | 0.549 | 1.000 | −2 |

*oxide fraction determined from the charge neutrality condition – equation (S3)

The stoichiometric coefficient $v_i$ of an element $i$ in the surface chemical formula was calculated from equation (S4).

$$v_i = \frac{x_i \sum_j f_j |z_j|}{\sum_i \sum_j x_i f_j |z_j|} \quad (S4)$$

The atomic fractions and normalised stoichiometric coefficients of the elements are listed in Table S3. The chemical formula was therefore determined as $PbSn_{0.34}Pb_{0.66}S_{2.0}$ (~$Pb_3SnSb_2S_6$). Oxygen, whose concentration is negligible, is omitted from the formula for the sake of simplicity.

The charge neutrality condition, equation (S3), has been found to be fully satisfied assuming only 23.6 % of the oxygen is present as metal oxides ($v_i$ values were calculated under this assumption). From this follows that the remaining 72.4 % of oxygen is bound as carbonaceous impurities. An indirect indication of oxygen being bound in multiple species is also evidenced by the O 1s peak (Fig. 3d of the main article). Broadness of this peak, centred around 532.3 eV, suggests that it is in fact made up of several oxygen-bound species with differing binding energies (metal-oxides: ~529 – 531 eV, organic oxygen: ~531 – 533 eV).[3]

**Table S3 | Elemental stoichiometric coefficients.**

| element | $v_i$ | 4 × $v_i$ | 12 × $v_i$ |
|---|---|---|---|
| Pb | 0.25 | 1.01 | 3.04 |
| Sn | 0.09 | 0.34 | 1.02 |
| Sb | 0.16 | 0.65 | 1.94 |
| O  | 0.01 | 0.03 | 0.10 |
| S  | 0.49 | 1.97 | 5.90 |

Considering previous reports of oxidation and carbonaceous contamination of other layered materials,[4, 5, 6, 7] it is useful to compare the surface composition of crystals that have been exposed to air for more than 24 h (aged surface) with those exposed immediately prior to the transfer to an XPS vacuum chamber (freshly cleaved surface). The XPS spectra obtained on aged and freshly cleaved franckeite surface are shown in Fig. S4a and S4b, respectively. The quantification analysis embedded in Fig. S4 suggests that the variations in element quantities are similar for both aged and freshly cleaved surfaces. As expected, there is an increase in the amount of surface carbon (by 13%) and oxygen (by 20%). Overall, it can be concluded that the differences between the aged and freshly cleaved surface are small and one would have to significantly decrease the time between the cleaving and XPS measurement in order to observe more pronounced changes.

Adventitious carbon and trace elements (Ag, Fe, Cl, and F) below 1 at%, which were observed intermittently, were not included in this analysis. It is reasonable to assume that the amounts of these elements will balance out, thus contributing a near-zero net charge to the chemical formula. They will therefore have little impact on the stoichiometric analysis described above. The differences between the EDXS and XPS quantification, in particular the relative concentrations of Sn and Fe, which are related to difference between bulk and surface of franckeite crystals, are discussed in the main text.

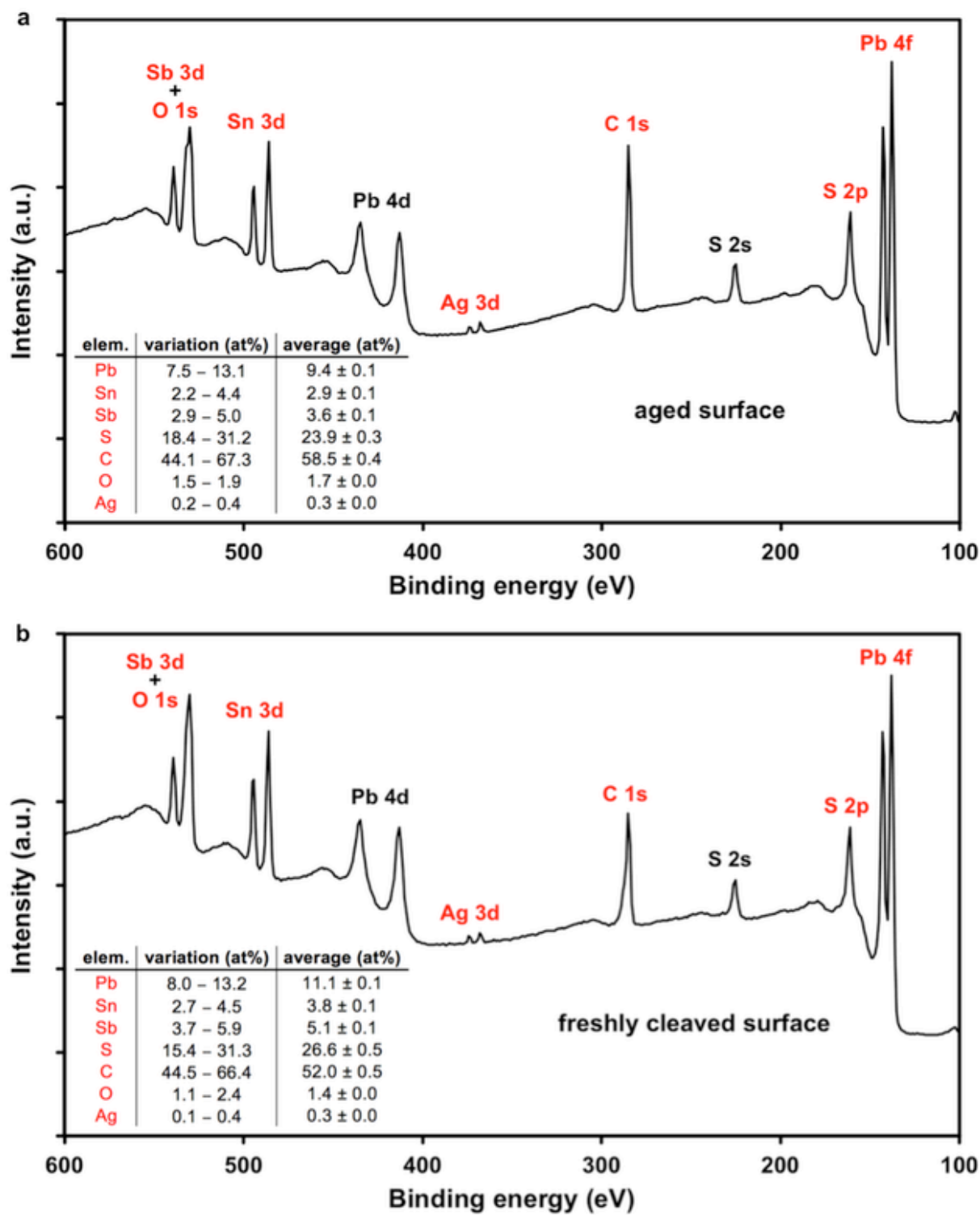

**Figure S4 | XPS of aged and freshly cleaved franckeite surface. a**, XPS spectrum of aged surface. **b**, XPS spectrum of freshly cleaved surface. Each spectrum is an average of 5 individual measurements from different parts of the crystal. The peaks used for quantification (shown the inset tables) are labelled in red, other major peaks are labelled in black. The spectra are normalised to the intensity of the Pb $4f_{7/2}$ peak (137.6 eV) with the adventitious carbon C 1s peak positioned at 284.7 eV.[8]

Fig. S5 displays the high-resolution XPS spectra of minority metallic elements, silver and iron, and adventitious surface carbon. Fig. S5a shows the Ag 3d doublet peak. Note that silver, as well as iron are common contaminants in galenite (PbS) and are therefore likely to follow more complex galenite-derived minerals, such as franckeite.[9, 10] Deconvolution of the Fe 2p doublet peak, superimposed on Sn $3p_{3/2}$ peak, is shown in Fig. S5b. The C 1s peak and its deconvolution to three components are shown in Fig. S5c. For the sake of simplicity and due to the large variation in the reported values of organic carbon binding energies, we have grouped the carbon-bound species into three categories, based on their approximate binding energy.[3] These are: 1) carbon-carbon $sp^2$ and $sp^3$ bonds, 2) carbon-hydrogen and carbon-oxygen (single) bonds, and 3) carbon-oxygen (double) and carbon-halogen bonds.

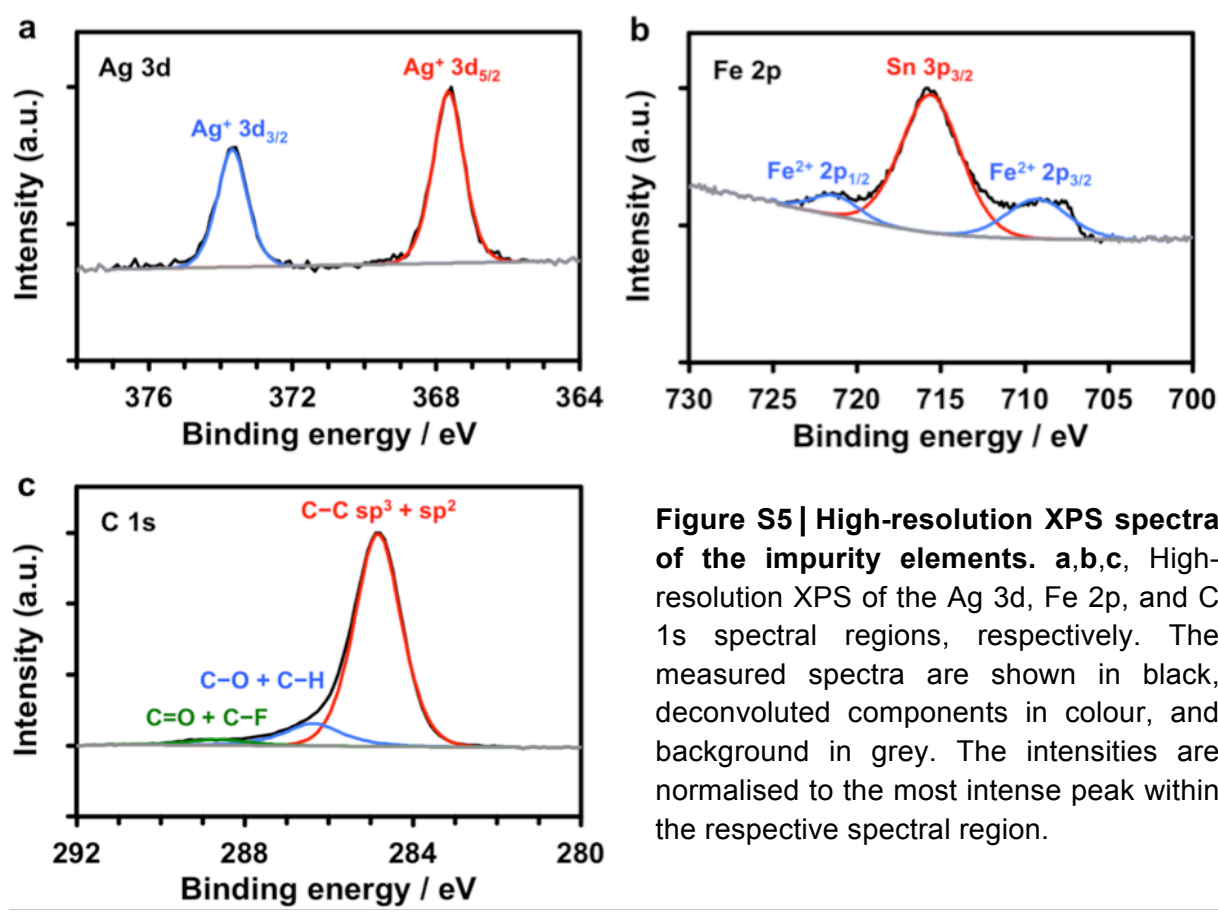

**Figure S5 | High-resolution XPS spectra of the impurity elements. a,b,c**, High-resolution XPS of the Ag 3d, Fe 2p, and C 1s spectral regions, respectively. The measured spectra are shown in black, deconvoluted components in colour, and background in grey. The intensities are normalised to the most intense peak within the respective spectral region.

**S4. Raman spectroscopy, optical microscopy, and atomic force microscopy**

Figure S6 shows the Raman spectra of bulk and monolayer franckeite crystals as well as the reference spectra of the underlying SiO$_2$ substrate. Qualitative differences in the bulk and substrate-corrected monolayer spectra are discussed in the main article. The uncorrected monolayer spectra in Fig. S6 show that the decrease in the crystal thickness is accompanied by appearance of a minor silicon peak, 2TA(X) peak at 302 cm$^{-1}$, and two major silicon peaks, F$_{2g}$ at 520 cm$^{-1}$ and 2TO(L) at 970 cm$^{-1}$ in the monolayer Raman spectra.[11] The intensities of these peaks originating from the underlying SiO$_2$ substrate are greatly reduced in bulk crystal (> 100 nm thickness). There is currently no Raman literature on franckeite and available databases only provide spectra recorded at high laser power densities,[12] which damage metal chalcogenides minerals as shown below (Figure S8).

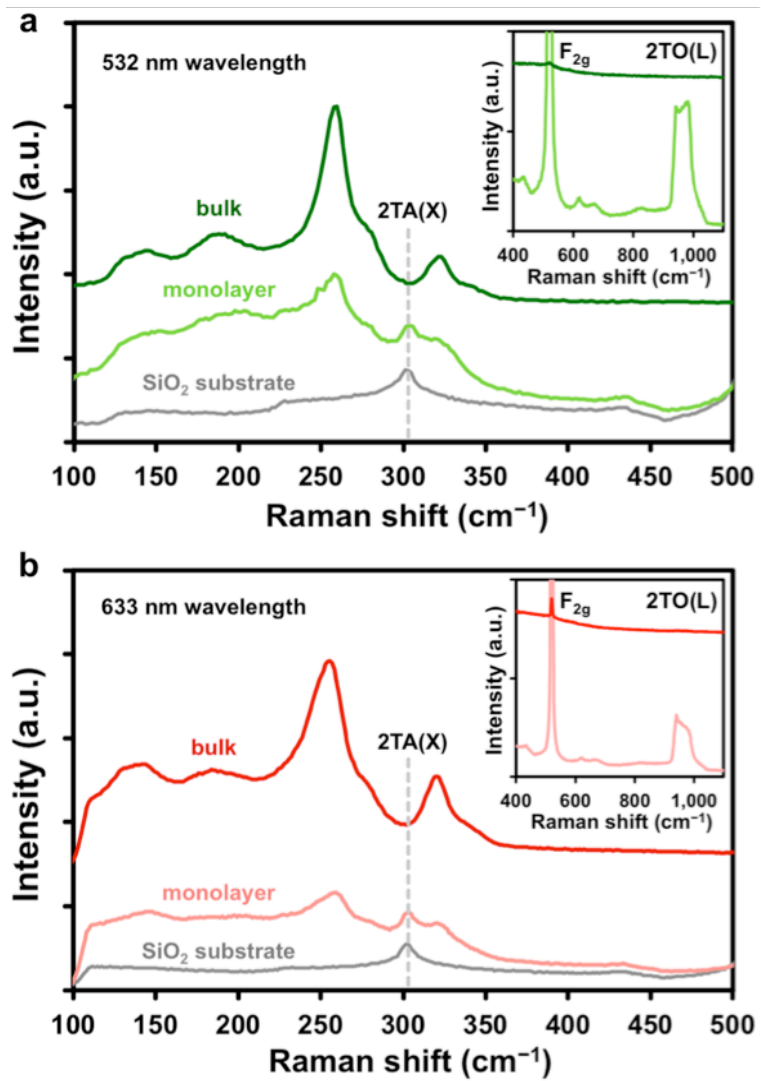

**Figure S6 | Raman spectra of monolayer and bulk franckeite crystals uncorrected for SiO$_2$ substrate. a**,**b**, Raman spectra of monolayer and bulk franckeite, and SiO$_2$ substrate, obtained using either a green laser of 532 nm wavelength (**a**) or a red laser of 633 nm wavelength (**b**). The insets show the spectral region encompassing the first- and second-order Raman Si bands. The intensities are normalised to the most intense bulk franckeite peak at 259 cm$^{-1}$ in **a** and 256 cm$^{-1}$ in **b**.

Further characterisation of typical needle-like few-layer franckeite crystals is shown in Figure S7. A BF optical micrograph in Fig. S7a shows such a crystal with an inset magnifying the monolayer/few-layer region. Fig. S7b shows a dark-field (DF) optical micrograph of the same crystal. Atomic force microscopy (AFM) image overlaid on the BF optical micrograph is shown in Fig. S7c. The zoom region in the inset in the top-right of the figure clearly shows the monolayer region, which was identified to have thickness of ~2.5 nm. Fig. S7d shows a Raman spectroscopy map overlaid on the BF optical micrograph. The intensity of the green colour corresponds to the intensity of the main Raman peak at 258.5 $cm^{-1}$, measured by the green laser (532 nm wavelength). The brightest colour corresponds to bulk crystals (thicker than 10 nm) and black colour corresponds to the $SiO_2$ substrate. Monolayer and few-layer regions are discernible despite their low signal level, clearly seen in Fig. S7e, which shows the Raman spectra of bulk, monolayer, and $SiO_2$ substrate extracted from the Raman map data at locations indicated by the coloured cross marks in Fig. S7d.

All the Raman spectra shown so far (both here and in the main article) were recorded at low laser power density, namely 0.03 MW $cm^{-2}$ for the green laser and 0.07 MW $cm^{-2}$ for the red laser. The crystals were also subjected to increasingly higher laser power density, results of which are shown in Figure S8. Fig. S8a and S8b are BF and DF optical micrographs of a crystal exposed to varied laser power densities of 0.03 MW $cm^{-2}$, 0.23 MW $cm^{-2}$, 0.42 MW $cm^{-2}$, and 2.45 MW $cm^{-2}$ (labelled 1 – 4 in Fig. S8a−b and S8f−g). Visible laser-induced damage of the crystals was observed for the high laser power densities ($\geq$ 0.42 MW $cm^{-2}$). Fig. S8c and S8d show BF and DF optical micrographs of a laser-irradiation pattern obtained at 0.42 MW $cm^{-2}$. The AFM image of this pattern in Fig. S8e reveals radial damage, matching the dimensions of the spherical laser spot size (~0.8 $\mu m^2$). Comparison between the laser-irradiation damage on bulk and monolayer crystals was also examined. The effect of the laser-induced damage on the Raman spectrum of a bulk crystal is visible in Fig. S8f. On the other hand, any changes in the Raman spectrum of a monolayer crystal (Fig. S8g) are indistinct due to its low intensity. The BF and DF optical micrographs of the bulk and monolayer crystal used to obtain this Raman spectrum are shown in Fig. S8h and S8i, respectively.

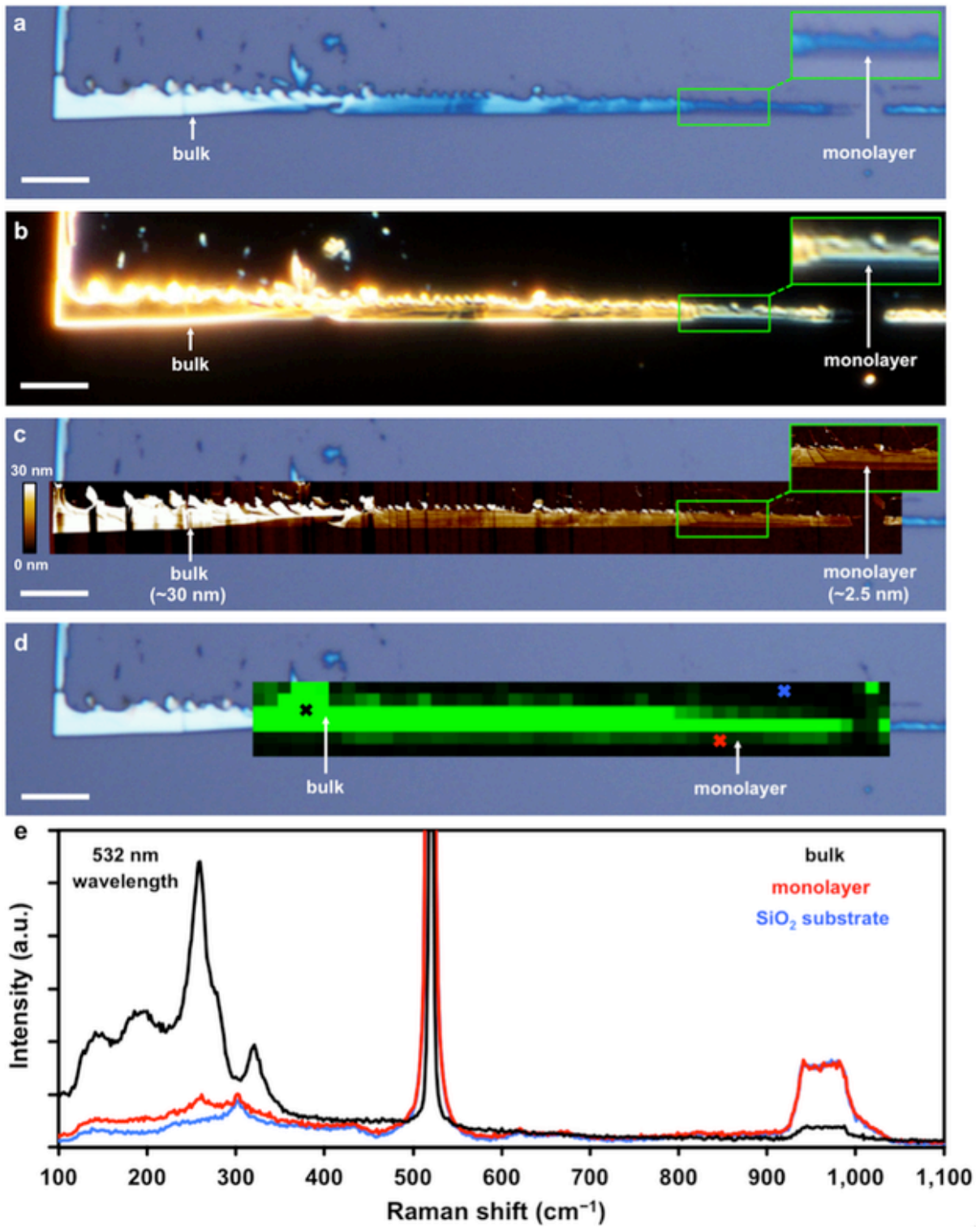

**Figure S7 | Optical microscopy, AFM, and Raman mapping of a thin franckeite crystal.**
**a**, BF optical micrograph the crystal exfoliated on SiO$_2$ substrate. **b**, DF optical micrograph of the same crystal. **c**, AFM image of a selected part of the crystal overlaid on the optical micrograph. The insets in **a–c** magnify the monolayer and few-layer region. **d**, Raman intensity map (at 258.5 cm$^{-1}$) of a selected part of the crystal overlaid on the optical micrograph. **e**, Raman spectra (using 532 nm wavelength) of the monolayer, bulk, and SiO$_2$ substrate, taken at locations indicated by coloured crosses in **d**. All scale bars denote 5 μm.

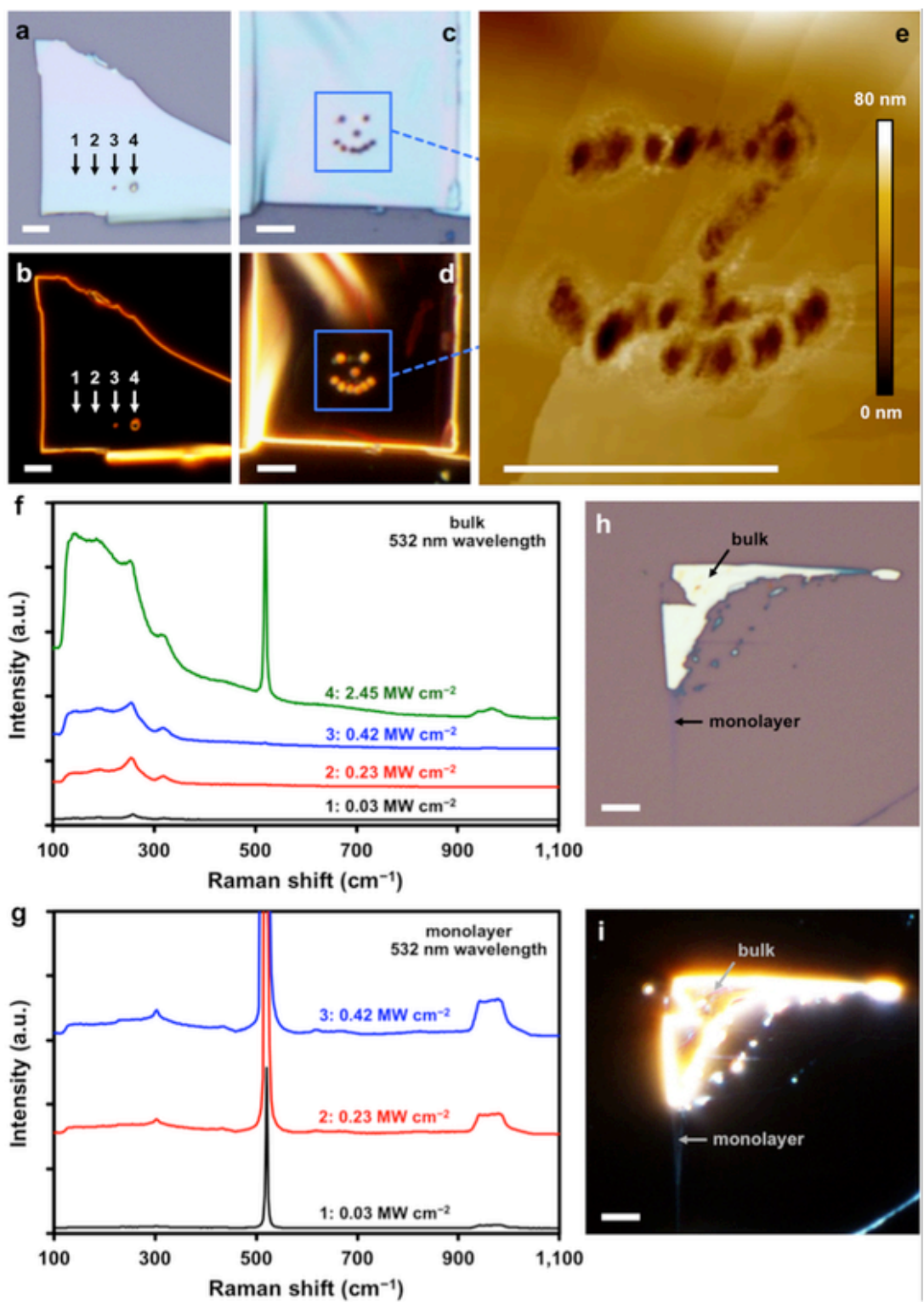

**Figure S8 | Laser irradiation of franckeite. a,b,** BF and DF optical micrographs of a bulk franckeite crystal subjected to increasingly higher laser irradiance for 30 s (1, 2, 3, and 4 correspond to 0.03, 0.23, 0.42, and 2.45 MW cm$^{-2}$). **c,d,** BF and DF optical micrographs of a laser-induced damage pattern using 0.42 MW cm$^{-2}$ irradiance for 30s per laser spot. **e,** AFM image of the same irradiation pattern. **f,g,** Raman spectra of bulk and monolayer crystals obtained using varied laser irradiance. **h,i,** BF and DF optical micrographs of the bulk and monolayer crystals used for Raman measurement. A 532 nm wavelength laser was used throughout. All scale bars denote 3 µm.

## S5. Electrochemical measurements

Cyclic voltammograms in pure electrolyte (capacitance measurement) obtained at varied scan rate and their dependence on consecutive scanning at constant scan rate are shown in Fig. S9a and S9b, respectively. Cyclic voltammograms with a redox mediator (electron transfer measurement) at varied scan rate is shown in Fig. S9c. Cyclic voltammograms in electrolyte with hydrochloric acid (hydrogen evolution measurement) are shown in Fig. S9d. The basal surface is very flat (root mean squared roughness, $R_q$, was typically $0.5 - 1.0$ nm) and therefore the geometric/active surface area effects are negligible. On the other hand, the uncertainty in geometric vs. active surface area on defects means that the electrochemical performance on these surfaces is more difficult to quantify.

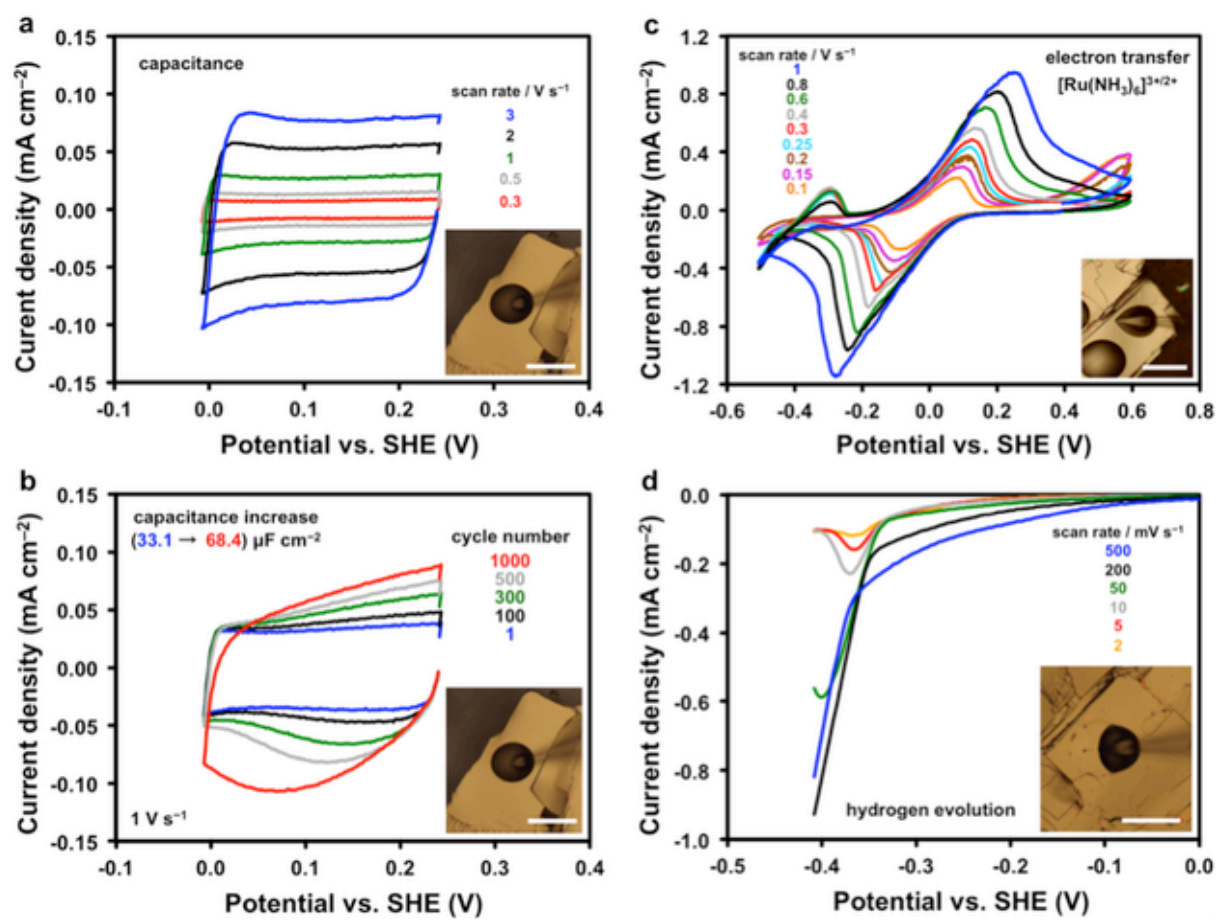

**Figure S9 | Capacitance, electron transfer, and hydrogen evolution measurements. a**, Capacitance measurement using cyclic voltammetry at varied scan rates (6 M LiCl). **b**, Capacitance dependence on consecutive voltammetric cycling. **c**, Electron transfer measurement using cyclic voltammetry at varied scan rate (with a $[Ru(NH_3)_6]^{3+/2+}$ redox mediator in 6 M LiCl). **d**, Hydrogen evolution measured by linear sweep voltammetry at varied scan rates (1 M HCl in 6 M LiCl). Optical micrographs of the liquid microdroplets used for the measurement are shown in the insets. All scale bars denote 50 μm.

The capacitance and electron transfer measurements were also examined at varied white light irradiance (illumination intensity) as shown in Figure S10. The illumination of the crystals was realised using the optical microscope objective (20×). The irradiance was calibrated using the 843-R power meter with a NIST-calibrated 818-Sl Si photodiode (Newport Spectra-Physics Ltd, UK). Fig. S10a, S10b, and S10c show the cyclic voltammograms in pure electrolyte, in electrolyte with $[Ru(NH_3)_6]^{3+/2+}$ redox mediator, and in electrolyte with $[Fe(CN)_6]^{3-/4-}$ redox mediator, respectively, obtained at constant scan rate and varied irradiance. Only small changes are observed with varying irradiance, which are likely to be related to thermal effects induced by illumination. This is in accordance with franckeite's narrow infrared band gap of 0.65 eV. The capacitance based on the data in Fig. S10b calculated from equation (3) slightly increases (by ~15 %) for the maximum irradiance in comparison to the zero-irradiance. The peak-to-peak separation of $[Ru(NH_3)_6]^{3+/2+}$ reduction/oxidation decreases by 8 mV for the maximum irradiance, indicating a slightly faster electron transfer rate (by ~14 %) in comparison to the zero-irradiance. Dependence of the $[Fe(CN)_6]^{3-/4-}$ reduction/oxidation peak separation on irradiance is complex, initially increasing and then decreasing slightly at high irradiance. It is possible that the inner-sphere $[Fe(CN)_6]^{3-/4-}$ mediator adsorbs at the crystalline surface, as observed previously on graphite,[4] which obscures the measurement. In a word, small changes in capacitance and electron transfer rate indicate franckeite is a degenerate, narrow band gap semiconductor.

The voltammetric background in pure supporting electrolyte over a large potential range was also examined at varied irradiance. Series of cyclic voltammograms in Fig. S10d show a redox activity in a wide potential region, which significantly increases with increasing irradiance. In order to deconvolute the contribution of the low and high potentials to the overall activity, separate voltammograms of the low and high potential regions were recorded (Fig. S10e and S10f). Analysis of the voltammograms indicates that the broad oxidative peak, centred around +0.5 V is associated with the reductive current at low potentials (below −0.35 V). Performing voltammetry in the low potential region (Fig. S10e) reveals a well-defined redox process, which becomes near-reversible at highest irradiance. When voltammetry is performed within the high potential region (Fig. S10f) the peak at +0.5 V disappears and the potential window is extended to ca. +0.8 V. The redox activity described above very likely contributes to the increased inherent capacitance within the narrow potential window (0 – 0.25 V vs. SHE), which therefore contains a pseudo-capacitive component. The redox activity originates from transitions between different oxidation states of the metals, although the exact peak assignment is difficult due to uncertainties in metal sulphides' standard redox potentials.[13]

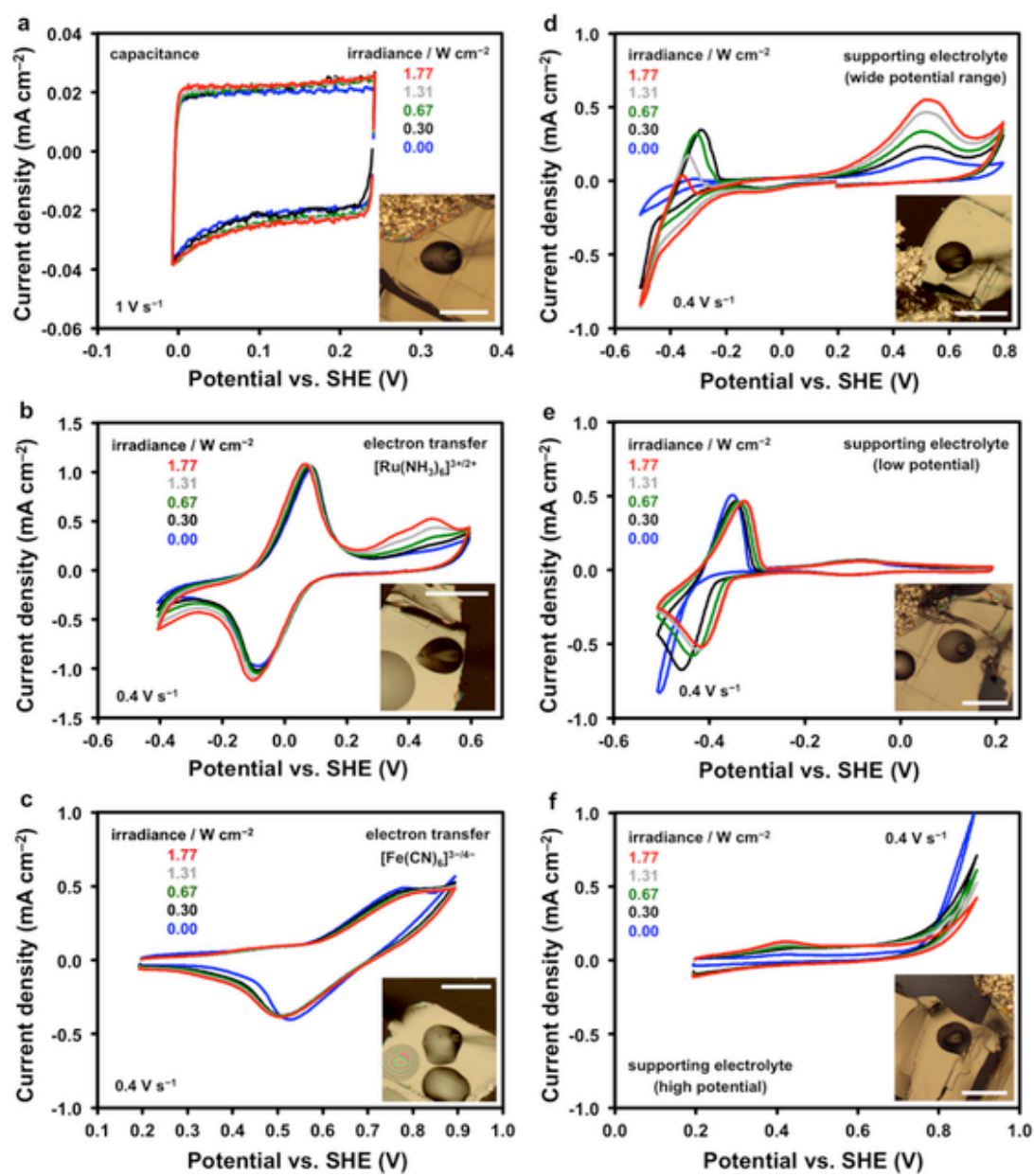

**Figure S10 | Voltammetry under varied irradiance. a,b,c**, Measurements of capacitance in 6 M LiCl, electron transfer using $[Ru(NH_3)_6]^{3+/2+}$, and electron transfer $[Fe(CN)_6]^{3-/4-}$ at varied irradiance (illumination intensity). **d,e,f**, Wide range, low, and high potential voltammograms in 6 M LiCl supporting electrolyte at varied irradiance. Optical micrographs of the liquid microdroplets on the franckeite surface, which were used for the measurement, are shown in the insets. All scale bars denote 50 μm.

**Supplementary References**

1. Wang S.& Kuo K. H. Crystal lattices and crystal chemistry of cylindrite and franckeite. *Acta Crystallogr., Sect. A: Found. Crystallogr.* **47**, 381-392 (1991).
2. Schaffer M., Schaffer B.& Ramasse Q. Sample preparation for atomic-resolution STEM at low voltages by FIB. *Ultramicroscopy* **114**, 62-71 (2012).
3. Benoit R. La Surface: XPS database; http://www.lasurface.com, accessed 30/04/2016. (ed^(eds). CNRS Orléans (2016).
4. Patel A. N.*, et al.* A new view of electrochemistry at highly oriented pyrolytic graphite. *J. Am. Chem. Soc.* **134**, 20117-20130 (2012).
5. Velický M.*, et al.* Electron Transfer Kinetics on Mono- and Multilayer Graphene. *ACS Nano* **8**, 10089-10100 (2014).
6. Velický M.*, et al.* Electron transfer kinetics on natural crystals of $MoS_2$ and graphite. *Phys. Chem. Chem. Phys.* **17**, 17844-17853 (2015).
7. Nioradze N., Chen R., Kurapati N., Khvataeva-Domanov A., Mabic S.& Amemiya S. Organic Contamination of Highly Oriented Pyrolytic Graphite As Studied by Scanning Electrochemical Microscopy. *Anal. Chem.* **87**, 4836-4843 (2015).
8. Barr T. L.& Seal S. Nature of the use of adventitious carbon as a binding energy standard. *J. Vac. Sci. Technol., A* **13**, 1239-1246 (1995).
9. Young C. A., Taylor P. R.& Anderson C. G. *Hydrometallurgy 2008: Proceedings of the Sixth International Symposium*. Society for Mining, Metallurgy, and Exploration (2008).
10. Moh G. H. Mutual $Pb^{2+}/Sn^{2+}$ substitution in sulfosalts. *Mineral. Petrol.* **36**, 191-204 (1987).
11. Wang R.-P.*, et al.* Raman spectral study of silicon nanowires: High-order scattering and phonon confinement effects. *Phys. Rev. B: Condens. Matter Mater. Phys.* **61**, 16827-16832 (2000).
12. Downs B. RRUFF Project: Mineral characterization database, http://www.rruff.info/franckeite/, accessed 26/04/2016. (ed^(eds). Department of Geosciences, University of Arizona (2016).
13. Licht S. Aqueous Solubilities, Solubility Products and Standard Oxidation-Reduction Potentials of the Metal Sulfides. *J. Electrochem. Soc.* **135**, 2971-2975 (1988).